\pdfoutput=1
\documentclass{pasa}%

\title[KIC\,7385478: An eclipsing binary with a $\gamma$ Doradus component]{KIC\,7385478: An eclipsing
binary with a $\gamma$ Doradus component}
\author[\"{O}zdarcan and Dal]{Orkun \"{O}zdarcan\thanks{Corresponding author: orkun.ozdarcan@ege.edu.tr},
Hasan Ali Dal\\
\affil{Ege University, Science Faculty, Department of Astronomy and Space Sciences, 35100 Bornova, 
\.{I}zmir, Turkey}}%
\jid{PASA}
\doi{10.1017/pas.\the\year.xxx}
\jyear{\the\year}

\usepackage[authoryear]{natbib}
\bibpunct{(}{)}{;}{a}{}{,}
\usepackage{aas_macros}
\usepackage{hyperref} 
\hypersetup{colorlinks,citecolor=blue,linkcolor=blue,urlcolor=blue}

\newcommand{\Msun}{\mbox{M$_{\odot}$}}
\newcommand{\Rsun}{\mbox{R$_{\odot}$}}

\newcommand{\kms}{\mbox{km\,s$^{-1}$}}

\def\na{NewA}

\begin{document}%
\begin{abstract}
We present spectroscopic and photometric analysis of the eclipsing binary KIC\,7385478.
We find that the system is formed by F1V + K4III-IV components. Combining results from analysis of
spectroscopic data and $Kepler$ photometry, we calculate masses and radii of the primary and the
secondary components as M$_{1}$ = 1.71 $\pm$ 0.08 \Msun, M$_{2}$ = 0.37 $\pm$ 0.04 \Msun~ and 
R$_{1}$ = 1.59 $\pm$ 0.03\Rsun, R$_{2}$ = 1.90 $\pm$ 0.03\Rsun, respectively. Position of the primary
component in HR diagram is in the region of $\gamma$ Doradus type pulsators and residuals from light
curve modeling exhibit additional light variation with a dominant period of $\sim$0.5 day. These are
clear evidences of the $\gamma$ Doradus type pulsations on the primary component. We also observe
occasional increase in amplitude of the residuals, where the orbital period becomes the most dominant
period. These may be attributed to the cool star activity originating from the secondary component.
\end{abstract}
\maketitle%
\section{Introduction}\label{S1}

High precision optical photometry from space telescopes is a milestone for modeling of 
eclipsing binary systems, where current models can not reproduce very small amplitude 
variations adequately. Even though, reasonable light curve modeling of these systems
provides not only absolute physical properties, but also reveals additional small 
amplitude light variations beside the eclipse events and binarity effects (i.e. reflection,
gravity darkening, etc...). This leads us to find accurate position of the components of 
these system on HR diagram and evaluate their variation nature more accurately.

There is a few indicators to reveal the internal layers of the stars. One of them is the stellar
pulsation. Unfortunately the pulsations can not be observable for each star. On the other hand, 
the initial analysis and some studies in the literature, such as \citet{Uyt11}, indicated that 
KIC\,7385478 is one of the candidates for the eclipsing binaries with pulsating component.
Moreover, the pulsation behaviour seen in the stars is very important pattern to understand the 
stellar itself and its evolution. According to the observations lasting as long as several decades
indicate that there are several type pulsating stars such as Cepheid, $\gamma$ Doradus, and 
$\delta$ Scuti type pulsating stars in the Instability Strip in the Hertzsprung-Russell diagram,
especially on the main sequence. All these types are separated by their locations in the Instability
Strip from each other. Analysing the pulsation frequencies, which is generally known as stellar
seismology called asteroseismology, the physical processes behind both the pulsating nature and 
stellar interiors can be revealed. This is why the pulsating stars have an important role to
understanding stellar evolution \citep{Cun07, Aer10}.

KIC\,7385478 is an interesting eclipsing binary whose out-of-eclipse variation
does not seem regular. The system was identified as a variable star in ASAS catalog 
(ASAS J195058+4259.8) for the first time \citep{2009AcA....59...33P} with a variation period 
of 1$^{d}$.6551 and $V - I$ color of 0$^{m}$.661. Variation amplitude in $V$ and $I$ filters 
are given as 0$^{m}$.12 and 0$^{m}$.19, respectively. Later, \citet{2011AJ....142..160S} 
compiled the $Kepler$ eclipsing binary catalog\footnote{http://keplerebs.villanova.edu/}, 
including KIC\,7385478. Extracted high precision light curves, ephemeris, 
period and effective temperature estimation for most of the system are provided in the catalog.
The catalog provides effective the temperature of KIC\,7385478 as 
6477 $K$. \citet{2012ApJS..199...30P} revised the temperature to 6735 $K$. More recently,
\citet{2014MNRAS.437.3473A} estimated the effective temperatures of the 
primary and the secondary components of KIC\,7385478 as 6346 $K$ and 4719 $K$, 
respectively, which are based on spectral energy distribution fitting. The ephemeris and the period
were revised by \citet{2016Borkovits_et_al}, who provided eclipse timing analysis of the system. Their 
analysis revealed the orbit of a third body physically bound to the eclipsing binary and they estimated
the minimum mass of the third body as 0.27\Msun. Beyond these studies, there is no comprehensive
analysis of the system published so far.

In this study, we focus on spectroscopic and photometric modeling of the system. In the next section
we give summary of photometric observations, and spectroscopic observations including reduction
process. Section~\ref{S3} comprises spectroscopic, orbital and light curve modeling together with
calculated physical properties and evolutionary status of the system. In addition we further give
analysis on residuals from light curve modeling in order to investigate out-of-eclipse light 
variations. In the final section, we summarize and discuss our findings.

\section{Observations and data reductions}\label{S2}

\subsection{Kepler photometry}\label{S2.1}

There is no ''$true$'' photometric filters in photometer of the $Kepler$ space telescope, hence $Kepler$ 
photometry contains no color information. However, response function of the photometer covers a very
broad wavelength range, which is between 4100 \AA~and 9100 \AA~and this allows to collect photons 
from a large part of the optical spectrum and increases the photometric precision, thus enables to
detect light variations with a sub-milimag amplitude, such as planet transit and small amplitude
oscillations. Photometric measurements are collected in short cadence and long cadence mode, which
have typical exposure times of 58.89 seconds and 29.4 minutes, respectively. In this study, we use 
long cadence (29.4 min) data of KIC\,7385478 available at $Kepler$ eclipsing binary catalog. 
We consider detrended and normalized fluxes \citep{2011AJ....141...83P, 2011AJ....142..160S} 
in the catalogue. However, observations from quarter 12 and 13 are not included in the
catalogue data file, hence we downloaded data files of these missing quarters from 
MAST\footnote{http://archive.stsci.edu/kepler/} data archive center, in {\it{fits}} format. 
In order to evaluate these data together with the catalogue data, we consider simple aperture
photometry (SAP) measurements in the {\it{fits}} data files of quarter 12 and 13. Then we detrend 
and normalize SAP fluxes as described in \cite{2011AJ....142..160S}. The final data set covers 
$\sim$4 years of time span with 65\,722 data points in total. MAST archive reports 0.9\% 
contamination level in the measurements, which indicates negligible contribution to the 
measured fluxes of KIC\,7385478, if any.

\subsection{Spectroscopy}\label{S2.2}

We carried out spectroscopic observations of KIC\,7385478 by 1.5 m Russian -- Turkish 
telescope equipped with Turkish Faint Object Spectrograph Camera (TFOSC) at Tubitak National
Observatory\footnote{http://www.tug.tubitak.gov.tr/rtt150\textunderscore tfosc.php}.
Spectra were recorded on a back illuminated 2048 $\times$ 2048 pixels CCD camera with a pixel size
of 15 $\times$ 15 $\mu m^{2}$. Observed spectra were obtained in \'echelle mode, which provides 
an effective resolution of R = $\lambda/\Delta\lambda$ = 2500 , which indicates $\Delta\lambda$ value of 2.6 \AA~around 6500 \AA~region. The spectra 
covers usable wavelength range between 3900 -- 9100 \AA~ in 11 orders.

We obtained nine optical spectra of KIC\,7385478 between 2014 and 2016 observing 
seasons. Exposure time of observations are 3200 s and 3600 s depending on atmospheric seeing 
conditions. Signal -- to -- noise ratio (SNR) of observed are between 80 and 145, which are
estimated via photon statistic . We further obtained high SNR spectra of 
$\iota$\,Psc (HD\,222368, F7V, $v_{r} = 5.656$ \kms), and 
HD\,184499 (G0V, $v_{r} = -166.1$ \kms) to use as radial velocity template and 
spectroscopic comparison. 

We used standard IRAF\footnote{The Image Reduction and Analysis Facility is hosted by the 
National Optical Astronomy Observatories in Tucson, Arizona at URL iraf.noao.edu.} packages and 
tasks to reduce all observations. In each observing run, several bias and halogen lamp (flat field) 
frames were obtained as well as comparison lamp (Fe-Ar) images, just before or after the target star 
observation. In the beginning of the reduction process, master bias frame obtained 
from nightly taken 8-10 bias frames were subtracted from all object, Fe-Ar comparison lamp and 
halogen lamp frames. Then, normalized master flat field frame was produced from bias corrected 
halogen lamp frames and all target and Fe-Ar spectra were divided by the normalized flat field 
frame. Cosmic rays removal and scattered light corrections were applied to the bias and flat field
corrected images. The spectra from reduced frames were extracted with IRAF task "$apall$" under
noao.imred.echelle package. Fe-Ar images were used for wavelength calibration and finally wavelength
calibrated target star spectra were normalized to the unity by using cubic spline functions.

\section{Analysis}\label{S3}

\subsection{Spectral type}\label{S3.1}

Instead of the published effective temperature estimations in the literature, we use 
the advantage of having medium resolution spectra of the system to estimate 
the spectral type and effective temperature, which would be more accurate.

\begin{figure}[!htb]
\centering
{\includegraphics[angle=0,scale=0.80,clip=true]{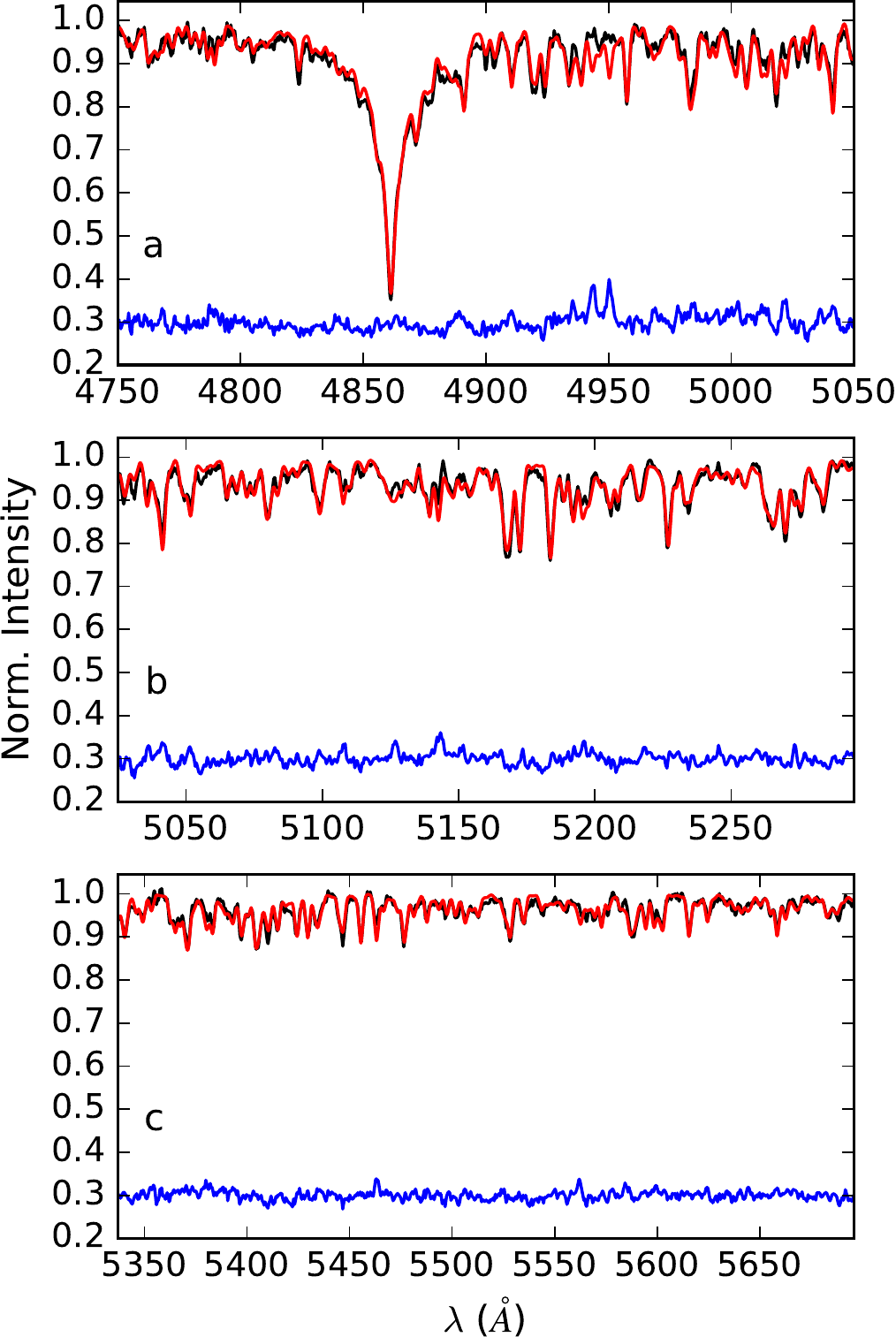}}
\caption{Representation of the observed (black), best matched (red) synthetic spectrum and 
residuals (blue) for three regions. Note that we shift the residuals upwards by 0.3 for the 
sake of simplicity. Panels $a$, $b$ and $c$ show the regions around H$_{\beta}$, Mg I triplet and 
metallic absorption lines around 5500 \AA, respectively.}
\label{F1}
\end{figure}

Our preliminary light curve analysis indicates that the secondary component makes $\sim$15\% 
contribution to the total light of the system, which means the contribution of the secondary
component in the spectra taken around secondary minimum would be negligible in our resolution.
We have a good spectrum taken at orbital phase $\sim$0.45 (see Table~\ref{T1})
where the signal of the secondary component fairly diminishes in the resolution of our spectrum,
so that we mainly observe the spectrum of the primary component. We adopt this spectrum as $reference~
spectrum$ for the primary component and compare it with the observed spectra of $\iota$\,Psc and 
HD\,184499. The comparison shows that the primary seems hotter than the both standard stars,
therefore, we switch to synthetic spectrum fitting method. In general, we use the latest version of
python framework $iSpec$ \citep{iSpec_Cuaresma_2014A&A}, which provides easy use of various synthetic
spectrum calculation codes. Among these codes, we adopt 
SPECTRUM\footnote{http://www.appstate.edu/$\sim$grayro/spectrum/spectrum.html} code 
\citep{1994AJ....107..742G} and calculate synthetic spectra by using ATLAS-9 
\citep{2004astro.ph..5087C} model atmospheres in conjunction with line list from the third version of
the Vienna atomic line database ($VALD3$) \citep{VALD3_2015PhyS...}. Grids of model atmospheres are
taken from the temperature range between 6000 K and 7500 K in steps of 250 K and gravity (log $g$)
range between 4.5 and 3.5. In all calculations, we adopt 2 \kms~of microturbulence velocity and solar
metallicity. Calculated spectra are convolved with a Gaussian line-spread function to match the
resolution of the reference spectrum and this convolution is done either by auxiliary program $smooth$ 
provided by the SPECTRUM code or by built-in function of $iSpec$ code. However, we do not consider 
the rotational broadening and other broadening mechanisms, which could be ignored due to the relatively
large instrumental broadening in our spectra. The models with the temperature 7000 K and log $g$
values 4.5 and 4.0 provide very close matches to the reference spectrum. However, calculated
physical properties of the system show that the log $g$ value of the primary component is 4.27 
(see Section~\ref{S3.4}). We also use synthetic spectrum fitting routine in $iSpec$,
which mainly adopts method of minimizing $\chi^{2}$ value of the difference between the reference
spectrum and synthetic spectra. For this purpose, we choose wavelength region of 4750 -- 5700 \AA~and
executed fitting routine by fixing log $g$ value and choosing different starting parameters for the 
temperature and metallicity. This method confirmed primary star temperature as 7000 K and
solar metallicity assumption. This temperature corresponds to the $F1$ spectral type according to
the calibration given by \citet{G05}. Considering calculated error of the temperature from $\chi^{2}$ 
minimization method, temperature steps in ATLAS-9 grids and current resolution of the observed
spectra, uncertainty of the temperature is estimated to be 200 K. We plot the observed reference 
spectrum and the best matched synthetic spectrum in Figure~\ref{F1}, for three different regions.

\subsection{Radial velocities and spectroscopic orbit}\label{S3.2}

\begin{figure}[!htb]
\centering
{\includegraphics[angle=0,scale=0.80,clip=true]{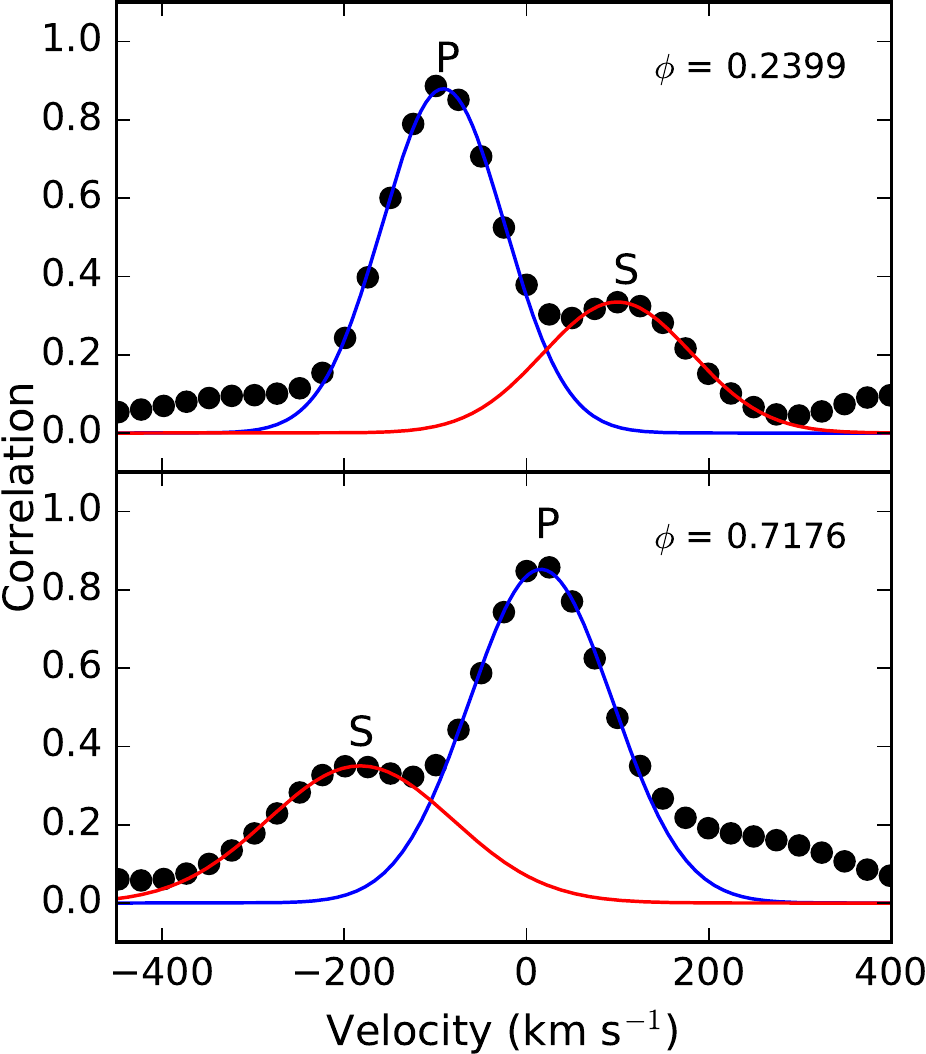}}
\caption{Cross-correlation functions of two spectra obtained in orbital quadratures. The letter $\phi$
shows corresponding orbital phase.}
\label{F2}
\end{figure}

We calculate radial velocities of the system by cross-correlating each observed spectrum with 
a template spectrum using $fxcor$ task in IRAF \citep{1979AJ.....84.1511T}. 
Here, we adopt $\iota$\,Psc as the template since it provides the most similar spectrum to 
KIC\,7385478 and obtained with the same instrumental setup. We consider absorption lines in 
\'echelle orders 3, 4, 5 and 6, which cover 5000 - 6800 \AA, except strongly blended lines and 
broad lines, such as H$_{\alpha}$ and Na I D lines. In Figure~\ref{F2}, we show cross-correlation 
functions of two spectra obtained around orbital quadratures. 

We give log of observations in Table~\ref{T1}, together with measured radial velocities and their 
standard errors. Orbital phases in the table are calculated via ephemeris and period given in
\citet{2016Borkovits_et_al}, which we also adopt for further analysis. Investigating Table~
\ref{T1}, one can easily notice the large standard error values for the measured radial velocities of 
secondary component. This is primarily caused by the small contribution of the secondary component to 
the total spectrum of the system, which can be evaluated from the cross correlation peaks of components 
given in Figure~\ref{F2}. We note that 3600 s of exposure time corresponds to 0.025 orbital phase step
and estimated radial velocity shift during this exposure time is about 8 \kms at the orbital quadratures 
for the secondary component, and much smaller for the primary component. These shifts are negligible
compared to the standard errors of the measured radial velocities, therefore we can safely ignore the
velocity shift due to the long exposure time.

\begin{table}
\setlength{\tabcolsep}{3pt}
\small
\caption{Log of spectroscopic observations together with measured radial velocities and their
corresponding standard errors ($\sigma$) in \kms.}\label{T1}
\begin{center}
\begin{tabular}{cccrrrr}
\hline\noalign{\smallskip}
      HJD    & Orbital  & Exposure & \multicolumn{2}{c}{Primary} &  \multicolumn{2}{c}{Secondary} \\
(24 00000+)  &  Phase   & time (s) & V$_{r}$ & $\sigma$ & V$_{r}$ & $\sigma$  \\
\hline\noalign{\smallskip}
 56845.4821  &  0.2399  &  3600   &  -54.5 & 5.2   &  159.0 & 19.9  \\
 56845.5248  &  0.2657  &  3600   &  -46.8 & 7.0   &  158.2 & 23.9  \\
 56846.4677  &  0.8352  &  3600   &   10.2 & 6.3   & -177.1 & 22.5  \\
 56887.4623  &  0.5984  &  3200   &    8.1 & 6.8   & -128.1 & 38.6  \\
 57591.5291  &  0.8948  &  3600   &   11.0 & 10.2  & -136.2 & 41.6  \\
 57600.5477  &  0.3425  &  3600   &  -46.8 & 10.9  &  132.6 & 32.4  \\
 57601.3798  &  0.8452  &  3600   &   24.3 & 13.2  & -152.0 & 38.1  \\
 57617.2875  &  0.4543  &  3600   &  -27.3 & 4.7   &   ---  & ---   \\
 57672.3540  &  0.7174  &  3600   &   24.4 & 8.4   & -191.5 & 25.3  \\
\noalign{\smallskip}\hline
\end{tabular}
\end{center}
\end{table}

\begin{figure}[!htb]
\centering
{\includegraphics[angle=0,scale=0.55,clip=true]{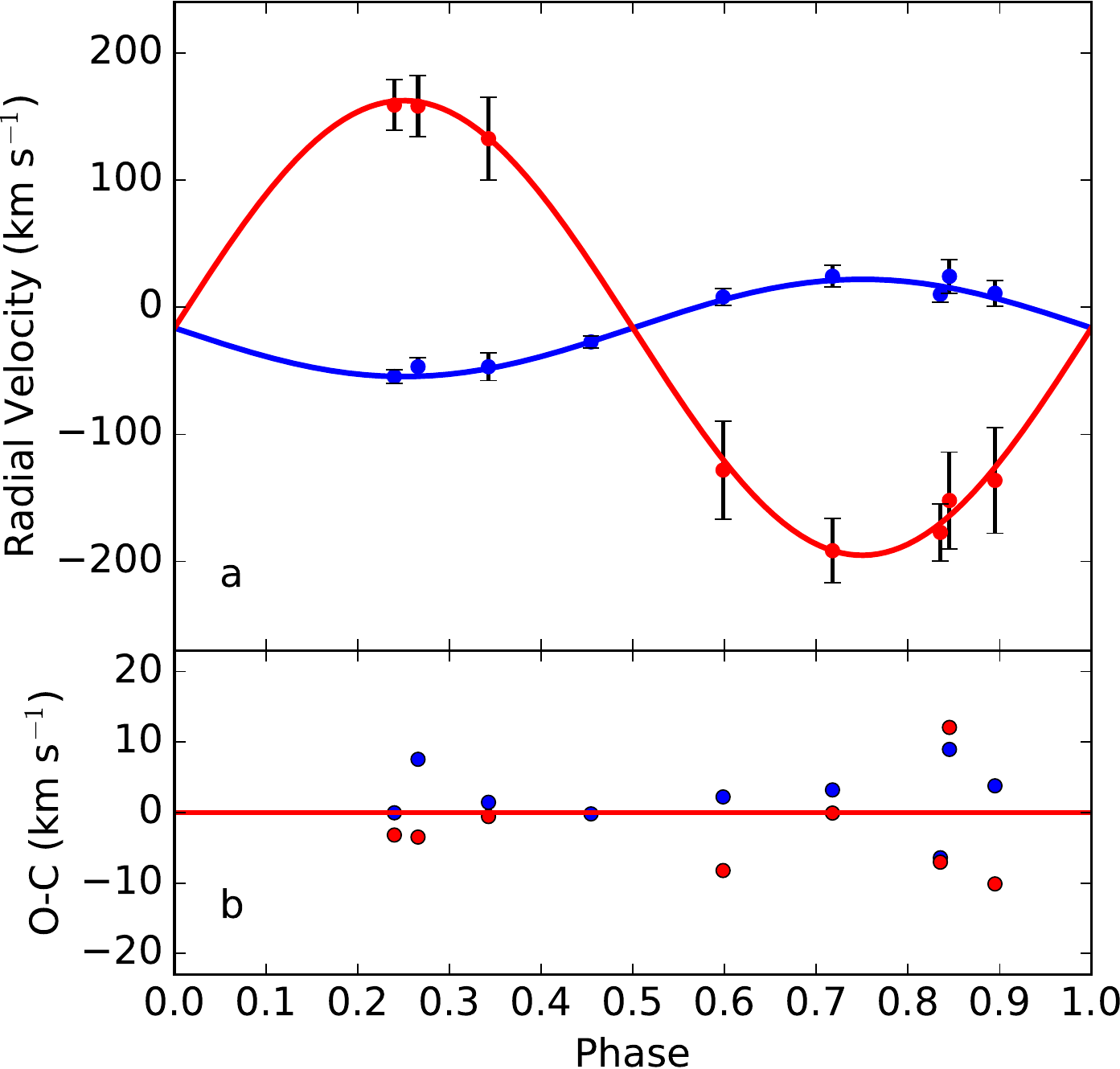}}
\caption{\textbf{a)} Observed radial velocities of the primary and the secondary (blue and red filled
circles, respectively) and their corresponding theoretical representations (blue and red curve).
\textbf{b)} Residuals from theoretical solution.}
\label{F3}
\end{figure}

Preliminary light curve examination of KIC\,7385478 indicates no evidence 
for eccentric orbit, therefore we calculate spectroscopic orbital elements of the system by 
assuming circular orbit. In addition, we fixed the orbital period and ephemeris to the values 
given by \citep{2016Borkovits_et_al}. We use a simple python script written by us, which applies
differential corrections by least squares method to all radial velocities, as described in
\cite{1935bist.book.....A}. We tabulate calculated spectroscopic orbital elements in Table~\ref{T2} 
and plot measured radial velocities together with theoretical spectroscopic orbit and residuals from
solution in Figure~\ref{F3}. Residuals, which belongs to the primary component scatters over 
zero level, while the residuals of the secondary component occupy the sub-zero level and this may be 
interpreted as if the whole fit could be improved further. However, if one consider the total rms of 
the fit given in Table~\ref{T2} (last row) as one $\sigma$, then the scatter of residuals is inside 
$\sim$1.5 $\sigma$ level and indicates that any further improvement to the orbital solution has no
statistical significance, but only cause slight changes in stellar parameters, which would
still stay inside corresponding standard error.

\begin{table}
\caption{Spectroscopic orbital elements of KIC\,7385478. $M{_1}$ and $M{_2}$
denote the masses of the primary and secondary component, respectively, while $M$ shows the total
mass of the system.}\label{T2}
\begin{center}
\begin{tabular}{cc}
\hline\noalign{\smallskip}
Parameter & Value \\
\hline\noalign{\smallskip}
$P_{\rm orb}$ (days)     &    1.655473 (fixed)  \\
$T_{\rm 0}$ (HJD2454+)   &  954.534784 (fixed)  \\
$\gamma$ (\kms)          &    -16.2$\pm$0.8     \\
$K_{1}$ (\kms)           &     38.3$\pm$2.7     \\
$K_{2}$ (\kms)           &    178.8$\pm$2.7     \\
$e$                      &      0 (fixed)       \\
$a\sin i$ (\Rsun)        &     7.10$\pm$0.12    \\
$M\sin^{3} i$ (\Msun)    &    1.754$\pm$0.061  \\
Mass ratio ($q=M{_2}/M{_1}$)         &     0.21$\pm$0.02    \\
fit rms (\kms )           &         6.5         \\
\noalign{\smallskip}\hline
\end{tabular}
\end{center}
\end{table}

\subsection{Light curve modeling}\label{S3.3}

\begin{figure}[!htb]
\centering
{\includegraphics[angle=0,scale=0.60,clip=true]{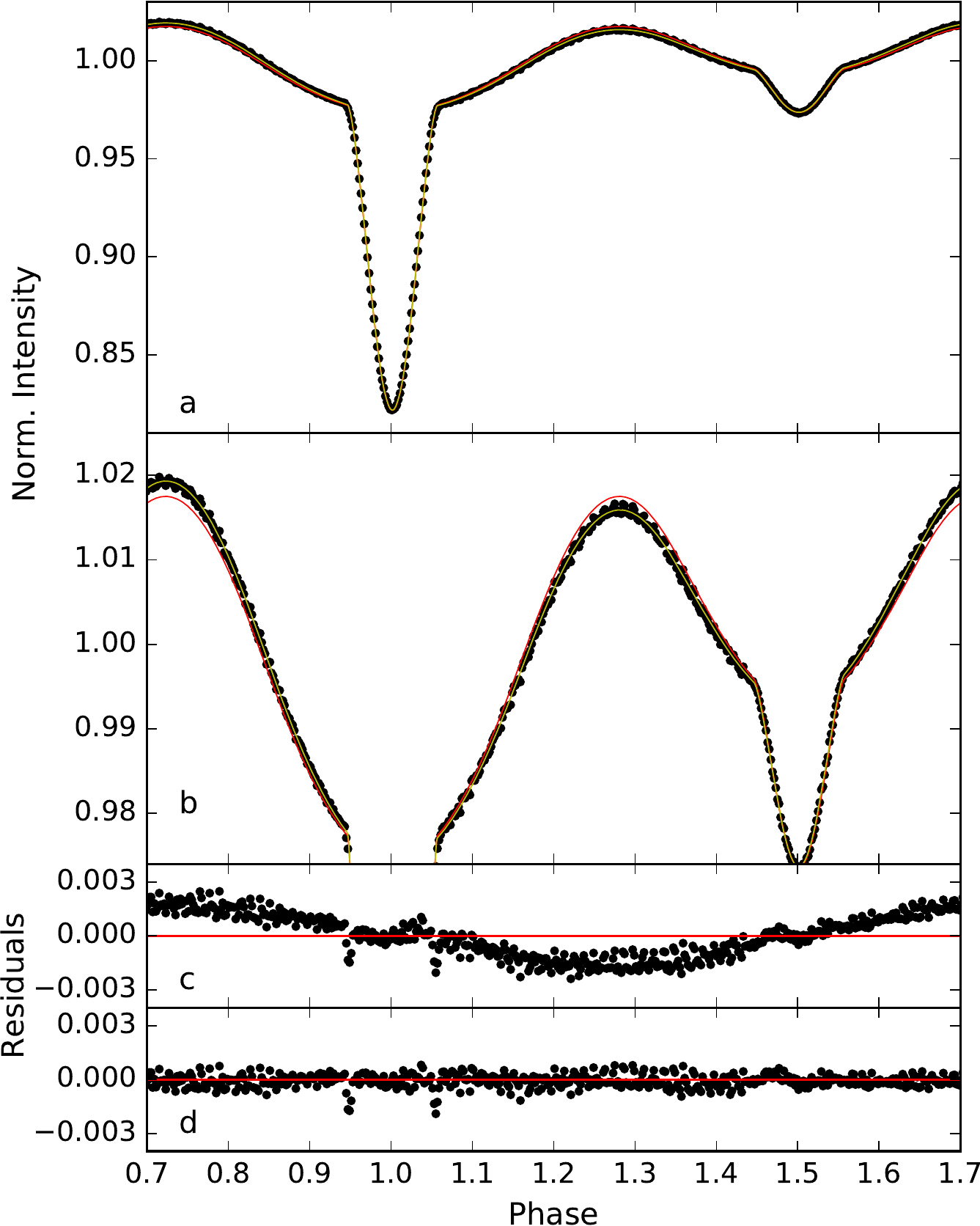}}
\caption{\textbf{a)} Phase binned light curve of KIC\,7385478 (black filled circles) 
together with best fit models with and without spot (yellow and red curves, respectively). 
\textbf{b)} Close up view of the light curve at light maxima. Panel \textbf{c} and 
\textbf{d} show residuals from the solutions without and with spot, respectively.}
\label{F4}
\end{figure}

In case of KIC\,7385478, light curve modeling would not be practical by considering 65\,722
long cadence data points, therefore we first phase the whole data with respect to the orbital period 
and then calculate binned light curve with a phase step of 0.002 by using freely available fortran code 
$lcbin$\footnote{http://www.astro.keele.ac.uk/$\sim$jkt/codes.html$\#$lcbin} 
written by John Southworth. We use the phase binned light curve for modeling the eclipsing binary. 
The first step of the light curve modeling is to find geometric and physical parameters of the system,
based on phase binned data and the second step is to calculate theoretical light curve with the best 
geometric and physical parameters and subtract the theoretical model from the whole long cadence data
in order to inspect possible out of eclipse variations. We plot the phase binned light curve 
in Figure~\ref{F4}, panel $a$.

By the quick inspection of the phase binned light curve, one may see that the system is on a circular 
orbit and there is a possibility of semi-detached configuration for the system. It may also be
noticed that light level around 0.75 phase is slightly higher than the light level around 0.25 
phase (see Figure~\ref{F4}, panel $b$), which means this effect is dominant through 4 years of 
$Kepler$ observations. Difference in the depth of the primary and secondary minima is another
important feature of the light curve, which indicates large difference between the temperatures of 
the components. Shape of the light curve at out of eclipse phases indicates variations beside the
geometric eclipse events, such as distortion in geometric shape of the components, reflection, spots,
and even oscillations as we will focus on in the next section.

We use 2015 version of the Wilson-Devinney code \citep{WD1971ApJ,WD2014ApJ} for light
curve modeling. Thanks to our spectroscopic observations, we have already determined the two most
critical parameters of the light curve modeling process, i.e. effective temperature of the primary
component and the mass ratio of the system. We fixed these two parameters during the modeling.
Since the effective temperature of the primary component is at critical location where the convective
outer envelope is very thin or almost becomes radiative, we carried out solutions by 
setting gravity darkening ($g_{1}$) and albedo ($A_{1}$) values to 0.32 and 0.5, respectively 
(for convective envelopes) and setting both parameters to 1.0 (for radiative envelopes). In both 
cases, there is almost negligible difference between solutions with radiative envelope and
convective envelope assumption, where convective envelope assumption leads to a slightly lower
residuals, thus we continue with 0.32 and 0.5 values for $g_{1}$ and $A_{1}$, respectively. We adopt
$g_{2}$ and $A_{2}$ values as 0.32 and 0.5 for the secondary component, which are typical for
convective envelopes. Both light curve and spectroscopic orbit solution indicate circular 
orbit for the system, therefore we assume synchronous rotation for the components, which is proper for
circular orbits, and fix rotation parameter ($F$) of each component to 1.0. Here, $F$ is defined as 
the ratio of the axial rotation rate to the orbital rate. Linear limb darkening coefficients ($x{_1},
x{_2}$) of the components are adopted from \cite{vanHamme1993AJ}.

We start analysis with detached configuration, therefore we consider inclination of the orbit 
($i$), temperature of the secondary component ($T_{2}$), dimensionless omega potentials of the primary 
and the secondary component ($\Omega_{1}$, $\Omega_{2}$), luminosity of the primary component 
($L_{1}$) as adjustable parameters. In addition, we observe a general phase shift of 0.002 
in the phase binned light curve, which possibly arises from a shift in the adopted ephemeris due to 
the third body \citep{2016Borkovits_et_al}, hence we also adjust phase shift during analysis. In a few
iterations, we observe that the $\Omega_{2}$ value jiggles around the inner critical potential value
and secondary component has entirely filled its Roche-lobe in the corresponding Roche geometry.
Then we switch to the semi-detached configuration and fixed $\Omega_{2}$. In this case, we consider
distorted shape of the secondary component and adopt $g_{2}$, $A_{2}$ and $x{_2}$ as adjustable 
parameters during iterations. After a few iterations, we achieve statistically the best parameter set,
however, we still observe that the residuals from the solution exhibit additional wave-like 
variation through an orbital cycle. 

During the iterations, we noticed that the contribution of the secondary component to the total 
light is not more than 15\%, therefore we expect negligible contribution from the secondary component 
to the wave-like variation in residuals. The most possible explanation could be that the Roche-lobe 
filled secondary star transfers its own mass to the primary component through the inner Lagrange point,
L1, of the system. The transferred mass possibly hits directly to the photosphere of the primary
component without forming an accretion disk, thus forms a local region warmer than the surrounding
photosphere. Therefore, we consider this hot region as a bright spot on the primary component and
add a single bright spot into our light curve model. At first step, we adopt spot longitude and 
radius as adjustable parameters together with eclipsing binary parameters and fixed the spot 
co-latitude and temperature factor to 45$^{\circ}$ and 1.03, respectively. After these parameters 
are adjusted, we fix the spot longitude and radius, and adjust spot co-latitude and temperature factor.
Until we achieve the best solution, we adjust different number of spot parameters simultaneously with a
different combinations. When we reach the best solution, we adopt all parameters (spot and eclipsing
binary) adjustable and run a single iteration in order to obtain statistical uncertainties. 
We tabulate our results in Table~\ref{T3}. Note that we do not give the internal error of the 
$T_{2}$ since it is unrealistically small ($\sim$ 1 K), therefore we adopt the uncertainty of $T_{1}$
estimated in Section~\ref{S3.1} as the uncertainty of $T_{2}$. In Figure~\ref{F4} we over plot the 
best fit models of spotless and spotted solutions (panel $a$ and $b$) and show residuals from both 
solutions in panel $c$ and $d$ in the same figure.


\begin{table}[!htb]
\caption{Light curve modeling results of KIC\,7385478. $\langle r_{1}\rangle$ and 
$\langle r_{2}\rangle$ denote mean fractional radii of the primary and the secondary components,
respectively. Spot parameters are given in the order of co-latitude ($\theta$), 
longitude ($\varphi$), radius ($r_{spot}$) and temperature factor ($TF$). 
Internal errors of the adjusted parameters are given in parentheses for the last digits.
Asterix symbols in the table denote fixed value for the corresponding parameters.}\label{T3}
\begin{center}
\begin{tabular}{cc}
\hline\noalign{\smallskip}
Parameter & Value \\
\hline\noalign{\smallskip}
$q$ &  0.21* \\
$T_{1}(K)$ &  7000* \\
$g_{1}$, $g_{2}$  & 0.32*, 0.270(5)\\
$A_{1}$, $A_{2}$  & 0.5*, 0.699(3)\\
$F_{1}$ = $F_{2}$  & 1.0* \\
phase shift        & 0.00160(2) \\
$i~(^{\circ})$ &  70.966(8)\\
$T_{2}(K)$ &  4293(150)\\
$\Omega_{1}$ & 4.9582(43)\\
$\Omega_{2}$ & 2.2574* \\
$L_{1}$/($L_{1}$+$L_{2})$ & 0.8590(6) \\
$x{_1bol},x{_2bol}$ & 0.471*, 0.531*\\
$x{_1}, x{_2}$  & 0.462*, 0.344(8) \\
$\langle r_{1}\rangle, \langle r_{2}\rangle$ & 0.2114(2), 0.2527* \\
$\theta~(^{\circ})$              &   98(6)   \\
$\varphi~(^{\circ})$            &   93(1) \\
$r_{spot}~(^{\circ})$    &   13(1)  \\
$TF$                 &   1.022(2) \\
Model rms           &     1.1 $\times$ 10$^{-4}$   \\
\noalign{\smallskip}\hline
\end{tabular}
\end{center}
\end{table}

\subsection{Physical properties and evolutionary status}\label{S3.4}

\begin{table}
\caption{Absolute physical properties of KIC\,7385478. Error of each parameter is 
given in paranthesis for the last digits.}\label{T4}
\begin{center}
\begin{tabular}{ccc}
\hline\noalign{\smallskip}
Parameter & Primary & Secondary \\
\hline\noalign{\smallskip}
Spectral Type      &  F1V     &  K4 III-IV \\
Mass (\Msun)       &  1.71(8) & 0.37(4) \\
Radius (\Rsun)     &  1.59(3) & 1.90(3) \\
Log $L/L_{\odot}$   &  0.737(52) & 0.043(82) \\
log $g$ (cgs)      &  4.269(7) & 3.444(31) \\
$M_{bol}$ (mag)      &  2.91(13) & 4.64(21) \\
\noalign{\smallskip}\hline
\end{tabular}
\end{center}
\end{table}

Combining results from spectroscopic orbital solution and light curve modeling, we calculate
absolute physical parameters of the system listed in Table~\ref{T4}. Inspecting the parameters,
we immediately see that the secondary component has a very large radius compared to its mass which
causes lower gravity compared to a typical main sequence star. A typical main sequence star 
has a radius of $\sim$0.41 \Rsun \citep{G05}. Adopting temperature calibration of
\citet{G05}, we estimate the spectral type of the secondary as K4 III-IV. All these indicate that 
the lower mass secondary component has already evolved off the main sequence, which seems 
contradictory to the stellar evolution theory. However, we know that the secondary component has 
already filled its Roche-lobe and there must be mass transfer from secondary to the primary via 
inner Lagrange point, L1. The situation could be explained as the 
secondary component was actually the more massive component in the system and as it evolved off 
the main sequence, it filled its Roche-lobe and started a mass transfer to the less massive component
(in our case, the primary component). We observe the effect of the mass transfer as hot spot on the
primary component. Continuous mass transfer in time finally reversed the mass ratio of the system,
therefore the final configuration became like a more massive main sequence star and a less massive 
sub-giant star. This is typical scenario adopted for Algols, which are well known to have 
reverted their mass ratio because of Roche lobe overflow.

\begin{figure}[!htb]
\centering
{\includegraphics[angle=0,scale=0.55,clip=true]{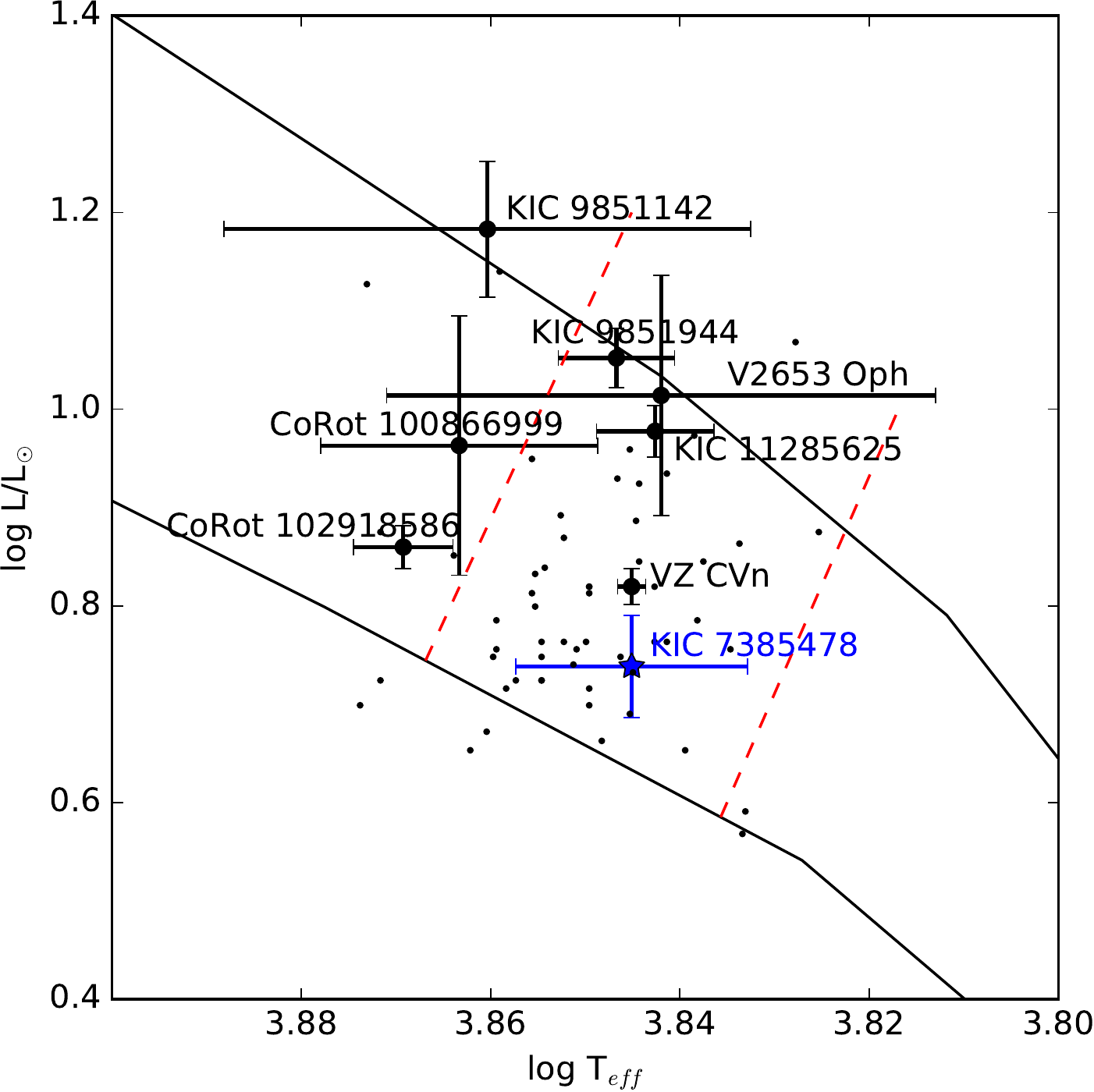}}
\caption{Position of the primary component of KIC\,7385478 on Hertzsprung-Russel diagram 
(star symbol in blue). Black dots show confirmed $\gamma$ Doradus stars from
\citet{2005Henry_gammadors}, filled large black circles denote discovered pulsating components in
eclipsing binaries. Red dashed lines indicate theoretical cool and hot boundary of $\gamma$ Doradus
instability strip \citep{2003Warner_inst_strip}. Black continuous curves show zero age and terminal 
age main sequences, taken from 
\cite{1998Pols_et_al}.}
\label{F5}
\end{figure}

\begin{figure*}[!htb]
\centering
{\includegraphics[angle=0,clip=true]{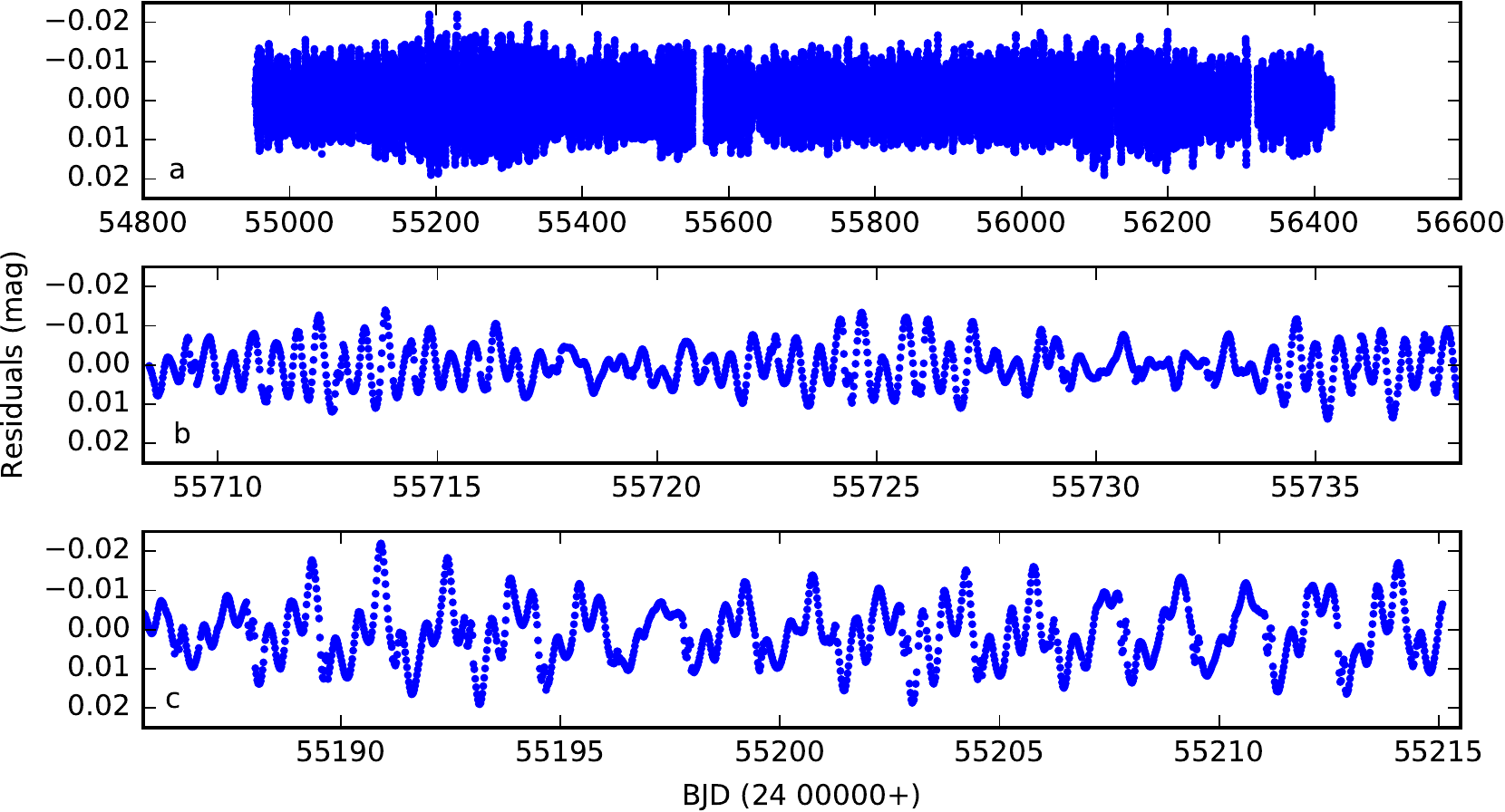}}
\caption{\textbf{a)} Residuals from whole long cadence data. \textbf{b)} Portion of residuals around
BJD = 2455723, where the variation amplitude is smaller. \textbf{c)} Residuals around BJD = 2455200, 
where the variation amplitude is larger.}
\label{F6}
\end{figure*}

In Figure~\ref{F5}, we show the position of the primary component of KIC\,7385478 on
Hertzsprung-Russel diagram, together with some eclipsing binaries with known $\gamma$ Doradus 
type pulsating components, i.e. VZ\,CVn \citep{2007_VZCVn_IB}, CoRot\,102918586
\citep{Corot102918586}, KIC\,11285625 \citep{KIC11285625_2013A&A}, KIC\,9851142
\citep{KIC9851142_2015NewA}, KIC\,9851944 \citep{KIC9851944_2016ApJ}, V2653\,Oph
\citep{V2653Oph_2016NewA}, CoRot\,100866999 \citep{CoRot100866999_2013A&A}. Our target is
located in the middle of the $\gamma$ Doradus instability strip given by 
\citet{2003Warner_inst_strip}, which takes us to the possibility of $\gamma$ Doradus type pulsation 
on the primary component.

\subsection{The out-of-eclipse variations}\label{S3.5}

In order to inspect the out-of-eclipse variations, we first construct a 
theoretical light curve by using the parameters in Table~\ref{T3}, then subtract it from whole 
long cadence data and obtain residuals. In this process, we first divide the whole long cadence 
data into subsets where each subset covers a single orbital cycle. Then, we run differential 
correction program of the Wilson-Devinney code by only adjusting ephemeris reference time of the
related subset and keeping all remaining parameters fixed. This process eliminates any shift in the
ephemeris reference time due to the light time travel effect caused by the third body and gives 
correct residuals.

In Figure~\ref{F6}, we plot the whole residuals (panel $a$), and a sample of residuals covering a time 
span of a month (panel $b$ and $c$). One may easily notice that removing eclipsing binary model from
the data unravels a clear variation with a dominant period of $\sim$ 0.5 day (panel $b$) with variable
amplitude, which suggests the possibility of $\gamma$ Doradus type pulsation 
\citep{gammadors_Kaye1999PASP} on the primary component, with a beat period of about 12 days. In
addition, we observe occasional increase in amplitude of residuals (panel $c$), where the most dominant
period becomes the orbital period, however by keeping 0.5 day variation as small humps and pits 
through an orbital cycle.

\begin{figure}[!htb]
\centering
{\includegraphics[angle=0,scale=0.65,clip=true]{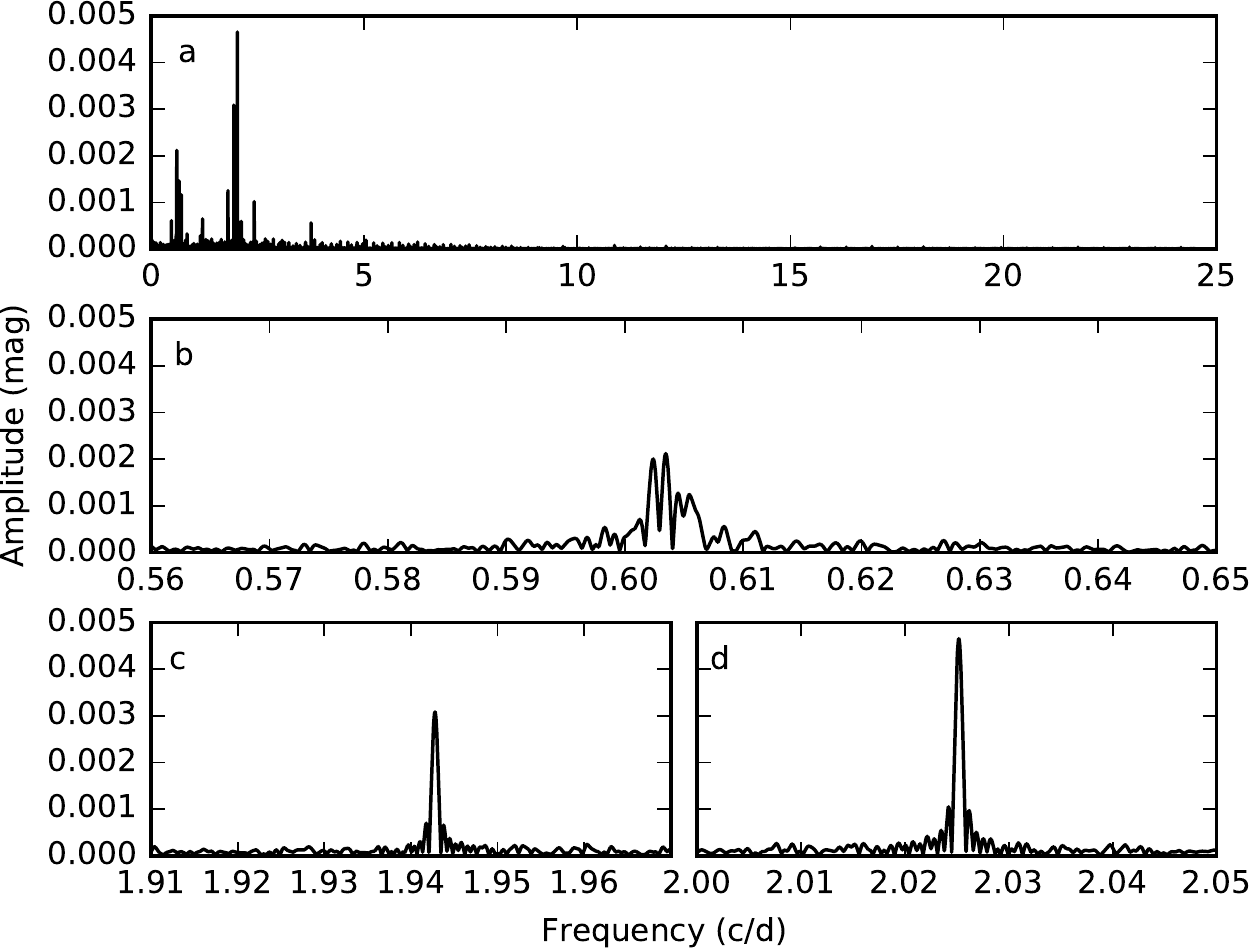}}
\caption{\textbf{a)} The whole amplitude spectrum of the residuals. \textbf{b)} Close view of the 
frequency range where the orbital frequency is located. Panel $c$ and $d$ are similar
to the panel $b$ but for pulsation frequencies.}
\label{F7}
\end{figure}

We apply multi-frequency analysis to the residual data using $pysca$ software package
\citep{pysca2014..301..421H} to investigate these variations. In case of continuous and long time
series photometry, $pysca$ is very practical to automatically extract significant frequencies above 
a defined SNR limit. We start with Fourier analysis of the data for the frequency between 0 and 24.498
cycle/day (c/d), where the 24.498 denotes the nyquist frequency. In Figure~\ref{F7}, we show amplitude
spectrum of the residuals (panel $a$). We observe that the dominant frequencies are located in lower
frequency region ($f$ $<$ 5 c/d) and consider this region for frequency extraction process. In this
process, the most dominant frequency in the amplitude spectrum is determined. Then its amplitude and 
phase are calculated via $A_{i} \Sigma sin[2\pi (f_{i}t + \phi_{i})]$, where t is the time of the 
corresponding measurement, while $A_{i}$, $f_{i}$ and $\phi_{i}$ show the amplitude, frequency and 
phase of the $i$th frequency, respectively. Next, this frequency is removed from the data and 
the same process is repeated for the remaining "\textit{prewhitened}" residuals. We adopt criteria of
\citet{Breger_1993}, which puts a lower SNR limit of 4 for a frequency to be accepted as significant.
Uncertainties of the extracted frequencies are estimated as $\sim$7$\times10^{-3}$ d$^{-1}$, which
is determined by using the Rayleigh criterion.

This process leads to 735 frequencies above SNR limit of 4. The most dominant two peaks are located 
at 2.0252 c/d and 1.9427 c/d, corresponding 0.4938 day 0.5147 day and we define these frequencies as 
$\gamma$ Doradus type pulsation frequencies. Third and fourth peaks are 0.6034 c/d and 0.6023 c/d,
corresponding 1.6573 day and 1.6603 day, and these frequencies indicate the orbital frequency. 
Beyond the first four frequencies, we do not find any independent frequency but low and high order
combination of pulsation and orbital frequencies. We list the extracted frequencies in Table~\ref{T_ap}.

\begin{figure}[!htb]
\centering
{\includegraphics[angle=0,scale=0.65,clip=true]{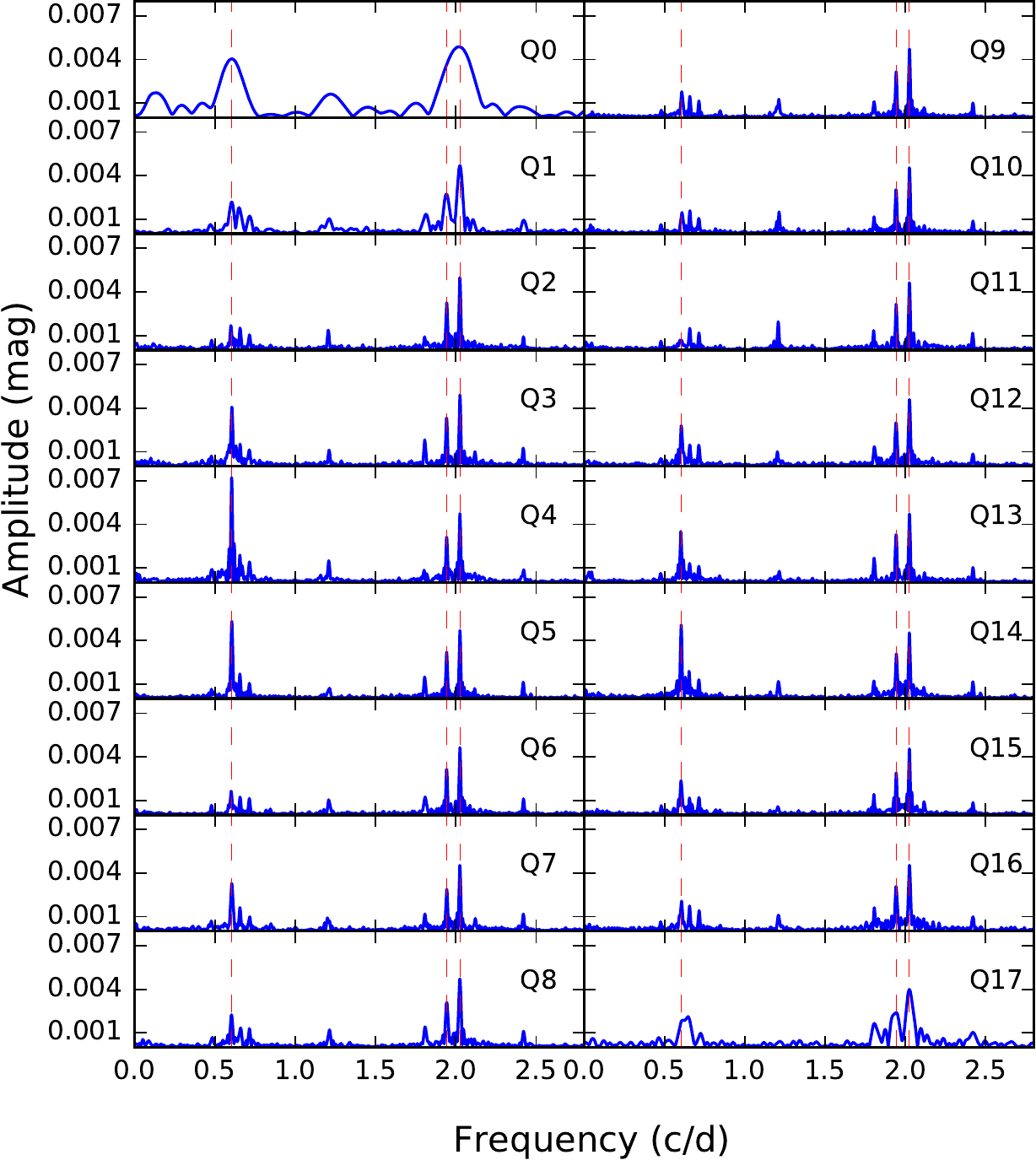}}
\caption{Amplitude spectrum of each quarter. We mark the location of the orbital frequency and the
most dominant two frequencies (i.e. pulsation frequencies) with red vertical dashed lines and label
each plot window according to its quarter number. Note that Q0 and Q17 contain less number of data 
points compared to the other quarters, thus peaks of dominant frequencies are broad in their amplitude
spectrum.}
\label{F8}
\end{figure}

In Figure~\ref{F7}, we also show close view around orbital frequency (panel $b$) and pulsation
frequencies (panel $c$ and $d$). One may easily notice that peaks at pulsation frequencies are single
and sharp peaks, while the peak at the orbital frequency is more shallow in amplitude and broader.
This picture leads us to calculate amplitude spectrum of long cadence data for each $Kepler$ data
quarter separately and investigate behavior of the dominant peaks in 4 years time range. We plot 
amplitude spectrum of each quarter separately in Figure~\ref{F8}. Amplitudes of the pulsation
frequencies are almost stable in all quarters, while the amplitude of the orbital frequency clearly
varies in time, which causes shallow and broader peak(s) at the orbital frequency in the amplitude
spectrum of the 4 year long cadence data.

\section{Summary and discussion}\label{S4}

Photometric and spectroscopic analysis of KIC\,7385478 shows that the system is a low
mass ratio ($q$ = 0.21) eclipsing binary formed by an F1V primary, and a K4III-IV secondary 
components which entirely fills its Roche lobe. This means the system can be classified as 
semi-detached. Physical properties of the components suggest that the low mass secondary
component has already evolved off the main sequence before the dwarf primary component, which is not
expected in the scope of the stellar evolution theory. Continuous mass transfer between the 
components could explain this situation, which could change the roles (and masses) of the components 
in long time scales. If the mass transfer does not lead to a disk, a hot spot forms on the 
photosphere of the mass gaining component, where the streaming matter directly hits to its
photosphere. Almost continuous $Kepler$ long cadence photometry through four years indicates a
brightness level difference between orbital quadrature phases (i.e. 0.25 and 0.75 phases), which 
could be explained by this kind of a hot spot on the primary (mass gaining) component.
At that point, the assumption of an impact system (no disc) can be justified by 
the mass ratio (0.22) and fractional radius of the primary (0.21) according to Lubow-Shu criterium 
\citep{1975ApJ...198..383L} for disc formation. Furthermore, we do not observe any emission feature 
in H$_{\alpha}$, indicates the absence of the disk and this is consistent with impact system 
assumption.

The position of the primary component on HR diagram corresponds to the middle of the region where 
$\gamma$ Doradus variables are located and suggests intrinsic light variations related to pulsation.
When we subtract the eclipsing binary model from the long cadence data, we see clear signal which has
a dominant period of $\sim$0.5 day indicating $\gamma$ Doradus type pulsations and confirms the 
intrinsic variation suggested by the position on HR diagram. However, we also observe occasional
increase in the amplitude of the residuals where the dominant period becomes the orbital period, 
including $\sim$0.5 day variation as a smaller amplitude variation. Multiple frequency analysis of
the residuals results in 735 frequencies, where the first two frequencies indicate $\gamma$ Doradus
type pulsations, while the third and the fourth frequencies correspond to the orbital frequency. 
Most of the remaining 731 frequencies correspond to either the orbital frequency or its harmonics.
This indicates an additional light variation with a period almost identical to the orbital one and 
this variation is usually suppressed by the light variation originating from pulsations.

Figure~\ref{F7} and Figure~\ref{F8} clearly show that the amplitude of the orbital frequency varies
in time and this causes broad and shallow peak structure at the orbital frequency in amplitude spectrum 
of 4 year data, which is another evidence for the additional light variation mentioned above. 
Possible star spot activity originating from the cool secondary component could easily cause this 
kind variation, thus leads to many low and high order combinations of the pulsation and orbital
frequencies in the amplitude spectrum, especially in period analysis of continuous long term data, 
which covers $\sim$4 years in case of KIC\,7385478. It is also known that stellar magnetic 
activity may cause orbital period modulations via mechanism proposed by \citet{1992ApJ...385..621A}.
Another possibility for orbital period modulations is variable mass transfer rate in the donor. Since
the typical time scales for Applegate mechanism and variable mass transfer rate are decades or longer, 
these are not comparable to the modulations observed in residual data of KIC\,7385478, therefore 
are not likely.

Estimated spectral type and luminosity class of the secondary component provide support for the
spot activity possibility. According to the eclipsing binary model, 14\% contribution of the secondary
component to the total light is expected at very broad band $Kepler$ filter, which means we may 
observe small amplitude light variation due to the possible spot activity of the secondary star.
However, when we check the observed spectra of the system, we do not observe any emission feature in 
Ca II H \& K lines, which are very sensitive to the chromospheric activity in cool stars. Considering
the temperatures and radii of the components, one may easily conclude 
that the contribution of the secondary component to the total light around 3950 \AA~ is almost
completely negligible, hence, considering only the spectroscopic indicators, we may not arrive at the
conclusion on the existence of star spot activity in case of KIC\,7385478. Therefore, 
we can still speculate that the secondary component might have star spot activity, which seems the 
most possible cause of the occasional amplitude increase in the light residuals.

\section*{Acknowledgments} 
We thank to TUBITAK for a partial support in using RTT150 (Russian-Turkish 1.5-m telescope in Antalya)
with project number 14BRTT150-667. This paper includes data collected by the Kepler mission. 
Funding for the Kepler mission is provided by the NASA Science Mission Directorate. Some of the data
presented in this paper were obtained from the Mikulski Archive for Space Telescopes (MAST). 
STScI is operated by the Association of Universities for Research in Astronomy, Inc., under NASA
contract NAS5-26555. Support for MAST for non-HST data is provided by the NASA Office of Space Science
via grant NNX13AC07G and by other grants and contracts.






\begin{appendix}

\begin{table*}\tiny
\section{Multi-frequency analysis results}
\caption{Extracted frequencies in multi-frequency analysis. N, F, A, P and SNR means
number, frequency (in c/d), amplitude (in mmag), phase and signal -- to --noise ratio,
respectively.}\label{T_ap}
\begin{center}
\resizebox{\textwidth}{!}{%
\begin{tabular}{cccccccccccccccccccc}
\hline\noalign{\smallskip}
N	&	F	&	A	&	P	&	SNR	\\
\hline\noalign{\smallskip}
1	&	2.0252	&	0.470	&	0.2618	&	275.8	\\
2	&	1.9428	&	0.315	&	0.6111	&	210.5	\\
3	&	0.6035	&	0.183	&	0.7206	&	105.5	\\
4	&	0.6023	&	0.169	&	0.5600	&	84.7	\\
5	&	0.6563	&	0.143	&	0.5643	&	76.2	\\
6	&	0.6055	&	0.127	&	0.1400	&	71.5	\\
7	&	1.8052	&	0.124	&	0.0554	&	91.7	\\
8	&	0.7138	&	0.115	&	0.7896	&	69.5	\\
9	&	2.4208	&	0.102	&	0.4827	&	87.2	\\
10	&	0.6048	&	0.090	&	0.3848	&	62.7	\\
11	&	0.6062	&	0.079	&	0.5564	&	50.9	\\
12	&	1.2087	&	0.058	&	0.1224	&	50.2	\\
13	&	1.2119	&	0.064	&	0.5436	&	48.1	\\
14	&	0.6008	&	0.065	&	0.7873	&	42.2	\\
15	&	0.4780	&	0.061	&	0.4596	&	42.2	\\
16	&	2.1178	&	0.060	&	0.4606	&	54.5	\\
17	&	3.7537	&	0.058	&	0.3023	&	67.9	\\
18	&	0.6110	&	0.062	&	0.5936	&	38.9	\\
19	&	0.6002	&	0.054	&	0.2745	&	39.7	\\
20	&	0.6084	&	0.045	&	0.1297	&	38.7	\\
21	&	0.6028	&	0.050	&	0.7349	&	36.8	\\
22	&	1.2041	&	0.051	&	0.3854	&	35.8	\\
23	&	0.5983	&	0.034	&	0.3999	&	29.4	\\
24	&	1.2051	&	0.036	&	0.5672	&	32.1	\\
25	&	0.8446	&	0.036	&	0.8880	&	27.6	\\
26	&	1.2110	&	0.036	&	0.6077	&	28.4	\\
27	&	1.8109	&	0.033	&	0.6449	&	29.9	\\
28	&	0.6041	&	0.035	&	0.0754	&	25.2	\\
29	&	0.6075	&	0.034	&	0.5545	&	26.5	\\
30	&	1.2099	&	0.032	&	0.5354	&	27.3	\\
31	&	0.5956	&	0.030	&	0.6500	&	22.3	\\
32	&	1.1569	&	0.026	&	0.9567	&	23.0	\\
33	&	0.5993	&	0.030	&	0.3687	&	23.2	\\
34	&	1.6469	&	0.024	&	0.7514	&	23.8	\\
35	&	0.5918	&	0.025	&	0.1310	&	19.6	\\
36	&	0.5970	&	0.021	&	0.9181	&	20.0	\\
37	&	0.0008	&	0.024	&	0.8344	&	18.6	\\
38	&	1.4211	&	0.022	&	0.8007	&	20.5	\\
39	&	2.8693	&	0.022	&	0.3247	&	24.7	\\
40	&	2.6815	&	0.022	&	0.0867	&	23.9	\\
41	&	0.6091	&	0.026	&	0.5645	&	18.4	\\
42	&	0.6101	&	0.022	&	0.8878	&	20.5	\\
43	&	1.2078	&	0.020	&	0.8653	&	20.3	\\
44	&	0.0129	&	0.021	&	0.2401	&	15.2	\\
45	&	1.8117	&	0.020	&	0.2413	&	20.2	\\
46	&	1.3322	&	0.019	&	0.7038	&	18.6	\\
47	&	0.6197	&	0.021	&	0.8756	&	16.6	\\
48	&	3.8374	&	0.020	&	0.9090	&	25.4	\\
49	&	1.1790	&	0.018	&	0.4249	&	17.4	\\
50	&	0.6662	&	0.017	&	0.4237	&	15.9	\\
51	&	3.0771	&	0.019	&	0.1901	&	22.3	\\
52	&	0.5902	&	0.016	&	0.0661	&	15.7	\\
53	&	0.6144	&	0.017	&	0.9196	&	15.4	\\
54	&	0.6162	&	0.016	&	0.5804	&	15.6	\\
55	&	1.2030	&	0.018	&	0.0640	&	17.3	\\
56	&	0.0441	&	0.018	&	0.9902	&	13.9	\\
57	&	0.0103	&	0.025	&	0.7827	&	13.7	\\
58	&	2.4687	&	0.018	&	0.5315	&	18.9	\\
59	&	4.4414	&	0.017	&	0.7160	&	24.4	\\
60	&	0.8171	&	0.018	&	0.0733	&	15.3	\\
61	&	2.7390	&	0.018	&	0.4742	&	19.9	\\
62	&	0.6216	&	0.018	&	0.4458	&	15.1	\\
63	&	1.2849	&	0.017	&	0.1149	&	16.3	\\
64	&	4.6194	&	0.016	&	0.5229	&	23.7	\\
65	&	2.3371	&	0.016	&	0.3393	&	17.2	\\
66	&	1.1995	&	0.018	&	0.3142	&	14.7	\\
67	&	1.2005	&	0.018	&	0.6592	&	17.3	\\
68	&	2.6293	&	0.014	&	0.3300	&	17.5	\\
69	&	0.4964	&	0.013	&	0.1267	&	13.2	\\
70	&	4.8330	&	0.013	&	0.6804	&	22.5	\\
71	&	3.0200	&	0.014	&	0.3884	&	18.4	\\
72	&	0.0309	&	0.014	&	0.8959	&	11.9	\\
73	&	1.2133	&	0.014	&	0.6766	&	14.5	\\
74	&	0.6301	&	0.011	&	0.8103	&	13.1	\\
75	&	3.1345	&	0.014	&	0.0059	&	18.4	\\
\end{tabular}
\begin{tabular}{cccccccccccccccccccc}
\hline\noalign{\smallskip}
N	&	F	&	A	&	P	&	SNR	\\
\hline\noalign{\smallskip}
76	&	0.0524	&	0.017	&	0.4613	&	11.4	\\
77	&	3.6250	&	0.010	&	0.0494	&	19.6	\\
78	&	1.8136	&	0.013	&	0.6465	&	15.2	\\
79	&	0.6280	&	0.014	&	0.5003	&	12.7	\\
80	&	0.6128	&	0.014	&	0.0515	&	13.1	\\
81	&	1.5403	&	0.014	&	0.3901	&	14.6	\\
82	&	4.0154	&	0.014	&	0.2136	&	20.2	\\
83	&	2.5991	&	0.014	&	0.2656	&	16.6	\\
84	&	0.6177	&	0.014	&	0.2119	&	12.7	\\
85	&	3.4113	&	0.013	&	0.2584	&	18.7	\\
86	&	0.0019	&	0.015	&	0.5863	&	11.6	\\
87	&	0.1168	&	0.012	&	0.4075	&	11.0	\\
88	&	3.2333	&	0.013	&	0.5021	&	18.2	\\
89	&	2.6566	&	0.014	&	0.8287	&	16.6	\\
90	&	1.3387	&	0.013	&	0.1568	&	14.1	\\
91	&	1.4590	&	0.013	&	0.0607	&	14.5	\\
92	&	1.2062	&	0.015	&	0.4634	&	14.7	\\
93	&	0.5925	&	0.011	&	0.3193	&	12.6	\\
94	&	1.7284	&	0.013	&	0.1816	&	14.9	\\
95	&	0.0826	&	0.009	&	0.3652	&	10.7	\\
96	&	0.6016	&	0.011	&	0.1482	&	12.8	\\
97	&	0.5811	&	0.013	&	0.7875	&	12.2	\\
98	&	2.1706	&	0.012	&	0.0829	&	15.3	\\
99	&	1.2105	&	0.013	&	0.4646	&	16.1	\\
100	&	2.4168	&	0.011	&	0.4929	&	15.5	\\
101	&	1.8152	&	0.016	&	0.2536	&	14.6	\\
102	&	1.2604	&	0.012	&	0.3536	&	13.4	\\
103	&	0.0039	&	0.010	&	0.0208	&	10.5	\\
104	&	2.2210	&	0.013	&	0.3379	&	15.1	\\
105	&	1.2201	&	0.011	&	0.5429	&	13.3	\\
106	&	2.8073	&	0.012	&	0.3263	&	16.0	\\
107	&	0.7347	&	0.012	&	0.5311	&	12.2	\\
108	&	0.0472	&	0.008	&	0.4934	&	10.1	\\
109	&	4.2289	&	0.010	&	0.0992	&	19.0	\\
110	&	0.2200	&	0.010	&	0.1354	&	10.8	\\
111	&	0.0095	&	0.014	&	0.4078	&	10.5	\\
112	&	3.7550	&	0.012	&	0.4206	&	17.9	\\
113	&	0.5854	&	0.014	&	0.8983	&	11.5	\\
114	&	0.2844	&	0.011	&	0.3917	&	10.5	\\
115	&	2.3913	&	0.011	&	0.1483	&	14.5	\\
116	&	1.0857	&	0.011	&	0.1677	&	12.5	\\
117	&	4.9631	&	0.011	&	0.7112	&	20.1	\\
118	&	0.5891	&	0.013	&	0.5727	&	11.8	\\
119	&	1.8169	&	0.011	&	0.4276	&	13.5	\\
120	&	1.3692	&	0.011	&	0.2257	&	12.6	\\
121	&	0.0432	&	0.014	&	0.1173	&	9.8	\\
122	&	1.4164	&	0.011	&	0.6597	&	12.3	\\
123	&	1.8143	&	0.012	&	0.6238	&	13.6	\\
124	&	0.0846	&	0.010	&	0.7156	&	9.7	\\
125	&	0.0069	&	0.012	&	0.4976	&	9.6	\\
126	&	2.9924	&	0.010	&	0.7486	&	14.9	\\
127	&	3.6239	&	0.012	&	0.3657	&	16.7	\\
128	&	3.0247	&	0.012	&	0.2936	&	15.0	\\
129	&	3.9680	&	0.011	&	0.6050	&	17.3	\\
130	&	4.3590	&	0.011	&	0.5049	&	18.0	\\
131	&	0.0565	&	0.013	&	0.4973	&	9.4	\\
132	&	2.4153	&	0.011	&	0.3813	&	13.8	\\
133	&	0.0533	&	0.011	&	0.3310	&	9.6	\\
134	&	0.5943	&	0.011	&	0.9709	&	10.9	\\
135	&	0.5597	&	0.011	&	0.2706	&	10.4	\\
136	&	2.6142	&	0.010	&	0.3255	&	14.1	\\
137	&	1.8084	&	0.010	&	0.6916	&	12.5	\\
138	&	4.7018	&	0.010	&	0.8318	&	17.8	\\
139	&	0.3418	&	0.008	&	0.7422	&	9.5	\\
140	&	0.0026	&	0.011	&	0.8688	&	9.1	\\
141	&	1.2124	&	0.012	&	0.7828	&	12.9	\\
142	&	4.8318	&	0.011	&	0.7306	&	18.1	\\
143	&	0.5695	&	0.008	&	0.4827	&	9.9	\\
144	&	0.6245	&	0.010	&	0.5124	&	10.1	\\
145	&	2.5468	&	0.009	&	0.2065	&	13.1	\\
146	&	0.1052	&	0.010	&	0.6602	&	9.0	\\
147	&	0.0336	&	0.010	&	0.5657	&	8.4	\\
148	&	0.1134	&	0.008	&	0.2306	&	9.0	\\
149	&	1.5992	&	0.009	&	0.8162	&	11.7	\\
150	&	0.0214	&	0.010	&	0.0713	&	8.6	\\
\end{tabular}
\begin{tabular}{cccccccccccccccccccc}
\hline\noalign{\smallskip}
N	&	F	&	A	&	P	&	SNR	\\
\hline\noalign{\smallskip}
151	&	0.0449	&	0.009	&	0.0795	&	9.2	\\
152	&	0.5877	&	0.011	&	0.1086	&	10.1	\\
153	&	0.5797	&	0.009	&	0.1969	&	10.0	\\
154	&	1.8645	&	0.009	&	0.9255	&	11.9	\\
155	&	1.2036	&	0.012	&	0.7640	&	12.9	\\
156	&	1.3178	&	0.009	&	0.8759	&	11.2	\\
157	&	0.5740	&	0.009	&	0.2844	&	9.9	\\
158	&	0.0288	&	0.007	&	0.6727	&	8.8	\\
159	&	4.2278	&	0.010	&	0.6653	&	16.5	\\
160	&	4.8369	&	0.009	&	0.8982	&	17.0	\\
161	&	1.7958	&	0.009	&	0.3709	&	11.6	\\
162	&	1.0052	&	0.009	&	0.6319	&	10.2	\\
163	&	0.9951	&	0.008	&	0.3070	&	10.3	\\
164	&	0.0895	&	0.012	&	0.7737	&	8.5	\\
165	&	0.0143	&	0.007	&	0.5967	&	8.2	\\
166	&	3.4937	&	0.009	&	0.4420	&	14.3	\\
167	&	3.6173	&	0.008	&	0.7603	&	14.6	\\
168	&	0.1378	&	0.010	&	0.3321	&	8.2	\\
169	&	0.0480	&	0.007	&	0.8814	&	8.3	\\
170	&	0.0186	&	0.006	&	0.1057	&	8.4	\\
171	&	0.0509	&	0.008	&	0.9190	&	8.6	\\
172	&	0.0803	&	0.010	&	0.6598	&	8.4	\\
173	&	0.0087	&	0.013	&	0.7656	&	8.7	\\
174	&	2.5259	&	0.008	&	0.5191	&	12.1	\\
175	&	2.3384	&	0.008	&	0.8083	&	11.9	\\
176	&	1.9985	&	0.009	&	0.7772	&	11.4	\\
177	&	4.4100	&	0.009	&	0.5855	&	15.4	\\
178	&	1.2272	&	0.009	&	0.9901	&	10.0	\\
179	&	0.5964	&	0.008	&	0.1089	&	9.3	\\
180	&	0.6187	&	0.012	&	0.3725	&	10.3	\\
181	&	0.6206	&	0.011	&	0.0193	&	10.2	\\
182	&	0.0315	&	0.014	&	0.7970	&	8.1	\\
183	&	0.0623	&	0.006	&	0.0421	&	8.4	\\
184	&	1.2171	&	0.010	&	0.5839	&	10.1	\\
185	&	1.1873	&	0.007	&	0.8383	&	10.1	\\
186	&	1.2160	&	0.009	&	0.6245	&	10.3	\\
187	&	0.0263	&	0.012	&	0.1299	&	8.4	\\
188	&	4.2328	&	0.010	&	0.9689	&	14.6	\\
189	&	4.9498	&	0.007	&	0.6767	&	15.7	\\
190	&	0.6314	&	0.004	&	0.0839	&	8.8	\\
191	&	1.9218	&	0.008	&	0.8825	&	10.8	\\
192	&	3.3306	&	0.008	&	0.2377	&	13.0	\\
193	&	3.6288	&	0.007	&	0.7552	&	13.6	\\
194	&	4.0977	&	0.008	&	0.3866	&	14.4	\\
195	&	0.2130	&	0.008	&	0.6613	&	8.0	\\
196	&	0.0049	&	0.010	&	0.9493	&	8.6	\\
197	&	0.0986	&	0.006	&	0.4770	&	7.8	\\
198	&	2.4135	&	0.006	&	0.6587	&	11.3	\\
199	&	4.8254	&	0.008	&	0.1253	&	15.5	\\
200	&	3.6223	&	0.007	&	0.8083	&	13.6	\\
201	&	1.2242	&	0.007	&	0.4654	&	9.5	\\
202	&	0.0368	&	0.007	&	0.6030	&	7.4	\\
203	&	0.5729	&	0.007	&	0.7046	&	8.7	\\
204	&	0.6579	&	0.007	&	0.6780	&	8.4	\\
205	&	0.1018	&	0.006	&	0.3971	&	7.5	\\
206	&	0.6838	&	0.008	&	0.5788	&	8.2	\\
207	&	0.1426	&	0.004	&	0.4217	&	7.3	\\
208	&	0.6423	&	0.010	&	0.2468	&	8.1	\\
209	&	1.8185	&	0.008	&	0.1807	&	9.9	\\
210	&	0.0864	&	0.010	&	0.2549	&	7.4	\\
211	&	0.3955	&	0.007	&	0.6943	&	7.7	\\
212	&	0.0670	&	0.007	&	0.2321	&	7.4	\\
213	&	3.0194	&	0.008	&	0.8203	&	11.5	\\
214	&	2.2032	&	0.007	&	0.1224	&	10.2	\\
215	&	0.0063	&	0.011	&	0.2367	&	7.7	\\
216	&	0.0630	&	0.008	&	0.2310	&	7.5	\\
217	&	0.0156	&	0.010	&	0.9619	&	7.5	\\
218	&	0.0198	&	0.009	&	0.9606	&	8.3	\\
219	&	0.6261	&	0.006	&	0.1145	&	8.1	\\
220	&	0.5823	&	0.010	&	0.3587	&	8.2	\\
221	&	2.4091	&	0.008	&	0.4653	&	10.3	\\
222	&	1.8695	&	0.007	&	0.6241	&	9.6	\\
223	&	2.4186	&	0.008	&	0.1687	&	10.4	\\
224	&	0.6488	&	0.007	&	0.7964	&	8.0	\\
225	&	0.0397	&	0.007	&	0.1003	&	7.3	\\
\end{tabular}
}
\end{center}
\end{table*}

\begin{table*}\tiny
\addtocounter{table}{-1}
\caption{Continued.}
\begin{center}
\resizebox{\textwidth}{!}{%
\begin{tabular}{cccccccccccccccccccc}
\hline\noalign{\smallskip}
N	&	F	&	A	&	P	&	SNR	\\
\hline\noalign{\smallskip}
226	&	0.0354	&	0.006	&	0.4586	&	7.6	\\
227	&	0.0938	&	0.007	&	0.1194	&	7.1	\\
228	&	1.1982	&	0.007	&	0.5649	&	8.6	\\
229	&	0.9672	&	0.006	&	0.0567	&	8.3	\\
230	&	2.8896	&	0.007	&	0.9090	&	10.8	\\
231	&	1.2020	&	0.009	&	0.0116	&	9.4	\\
232	&	3.1509	&	0.007	&	0.9775	&	11.2	\\
233	&	0.5869	&	0.009	&	0.8703	&	8.1	\\
234	&	1.2355	&	0.007	&	0.5330	&	8.6	\\
235	&	3.0208	&	0.007	&	0.4495	&	11.1	\\
236	&	2.7926	&	0.006	&	0.8003	&	10.5	\\
237	&	0.3717	&	0.007	&	0.5017	&	7.2	\\
238	&	1.2140	&	0.008	&	0.0111	&	8.7	\\
239	&	0.0222	&	0.009	&	0.8167	&	6.8	\\
240	&	0.5759	&	0.007	&	0.5043	&	7.7	\\
241	&	0.5772	&	0.006	&	0.1883	&	8.2	\\
242	&	0.0739	&	0.006	&	0.3209	&	6.7	\\
243	&	0.0574	&	0.005	&	0.2890	&	7.2	\\
244	&	0.0077	&	0.006	&	0.9595	&	7.0	\\
245	&	4.4674	&	0.006	&	0.2380	&	12.8	\\
246	&	0.9993	&	0.006	&	0.4145	&	8.1	\\
247	&	0.0322	&	0.007	&	0.6979	&	6.5	\\
248	&	1.1968	&	0.006	&	0.3491	&	8.2	\\
249	&	0.6349	&	0.004	&	0.1773	&	7.5	\\
250	&	0.0787	&	0.006	&	0.2070	&	6.8	\\
251	&	4.6394	&	0.006	&	0.1738	&	12.8	\\
252	&	2.0074	&	0.006	&	0.1696	&	9.1	\\
253	&	0.6599	&	0.006	&	0.8010	&	7.3	\\
254	&	1.2186	&	0.006	&	0.7455	&	8.1	\\
255	&	2.4178	&	0.007	&	0.2341	&	10.0	\\
256	&	0.1532	&	0.006	&	0.5155	&	6.4	\\
257	&	3.6261	&	0.008	&	0.0635	&	11.5	\\
258	&	1.8128	&	0.006	&	0.4256	&	9.2	\\
259	&	0.0178	&	0.009	&	0.9468	&	7.1	\\
260	&	0.1111	&	0.007	&	0.2124	&	6.6	\\
261	&	0.0609	&	0.009	&	0.2394	&	6.6	\\
262	&	0.1258	&	0.009	&	0.3570	&	6.6	\\
263	&	1.2226	&	0.008	&	0.0595	&	8.1	\\
264	&	0.6411	&	0.006	&	0.8622	&	7.2	\\
265	&	0.6323	&	0.008	&	0.4660	&	7.4	\\
266	&	3.0235	&	0.007	&	0.6443	&	10.0	\\
267	&	4.8395	&	0.006	&	0.7207	&	12.3	\\
268	&	3.0726	&	0.005	&	0.2607	&	9.9	\\
269	&	0.0494	&	0.007	&	0.2219	&	6.7	\\
270	&	0.0755	&	0.009	&	0.6666	&	6.8	\\
271	&	0.5363	&	0.004	&	0.8838	&	6.9	\\
272	&	1.0775	&	0.006	&	0.5053	&	7.6	\\
273	&	4.2214	&	0.006	&	0.7295	&	11.4	\\
274	&	1.9485	&	0.006	&	0.3600	&	8.5	\\
275	&	2.0246	&	0.007	&	0.7458	&	8.6	\\
276	&	0.5568	&	0.007	&	0.1948	&	6.7	\\
277	&	0.5676	&	0.005	&	0.3004	&	6.8	\\
278	&	0.4949	&	0.006	&	0.3021	&	6.7	\\
279	&	4.8282	&	0.006	&	0.0358	&	11.8	\\
280	&	4.8343	&	0.007	&	0.7484	&	12.1	\\
281	&	0.1558	&	0.006	&	0.2316	&	6.2	\\
282	&	0.1548	&	0.007	&	0.8164	&	7.1	\\
283	&	0.1661	&	0.006	&	0.3987	&	6.2	\\
284	&	0.5587	&	0.006	&	0.9196	&	7.1	\\
285	&	0.0972	&	0.005	&	0.5152	&	6.0	\\
286	&	0.0644	&	0.005	&	0.9109	&	6.2	\\
287	&	1.9416	&	0.006	&	0.6954	&	8.5	\\
288	&	1.1844	&	0.006	&	0.0778	&	7.4	\\
289	&	1.2148	&	0.007	&	0.6590	&	7.7	\\
290	&	2.4615	&	0.005	&	0.7649	&	8.7	\\
291	&	1.1953	&	0.005	&	0.1796	&	7.5	\\
292	&	0.1286	&	0.006	&	0.4343	&	6.2	\\
293	&	2.4120	&	0.006	&	0.6780	&	8.6	\\
294	&	2.4219	&	0.006	&	0.3595	&	9.1	\\
295	&	0.6379	&	0.004	&	0.7334	&	6.8	\\
296	&	4.2270	&	0.006	&	0.2032	&	10.9	\\
297	&	0.1249	&	0.005	&	0.5003	&	6.2	\\
298	&	4.8304	&	0.005	&	0.5935	&	11.5	\\
299	&	3.6203	&	0.006	&	0.6814	&	10.0	\\
300	&	1.3331	&	0.006	&	0.5969	&	7.6	\\
\end{tabular}

\begin{tabular}{cccccccccccccccccccc}
\hline\noalign{\smallskip}
N	&	F	&	A	&	P	&	SNR	\\
\hline\noalign{\smallskip}
301	&	0.1817	&	0.005	&	0.4327	&	6.0	\\
302	&	2.0261	&	0.005	&	0.5292	&	8.5	\\
303	&	3.6231	&	0.005	&	0.0308	&	10.3	\\
304	&	0.6272	&	0.004	&	0.4733	&	6.9	\\
305	&	0.5521	&	0.006	&	0.5181	&	6.4	\\
306	&	0.3340	&	0.004	&	0.7556	&	6.1	\\
307	&	1.8001	&	0.005	&	0.0146	&	7.8	\\
308	&	4.8848	&	0.005	&	0.8176	&	11.1	\\
309	&	3.6273	&	0.006	&	0.9809	&	10.2	\\
310	&	0.6629	&	0.006	&	0.2632	&	6.5	\\
311	&	0.6367	&	0.004	&	0.9079	&	7.0	\\
312	&	0.5779	&	0.004	&	0.0337	&	6.7	\\
313	&	0.5818	&	0.006	&	0.3530	&	7.0	\\
314	&	2.5455	&	0.005	&	0.4428	&	8.3	\\
315	&	2.6639	&	0.005	&	0.3763	&	8.5	\\
316	&	0.5716	&	0.005	&	0.1861	&	6.5	\\
317	&	0.1264	&	0.005	&	0.8625	&	5.8	\\
318	&	0.1182	&	0.005	&	0.3956	&	5.9	\\
319	&	3.6767	&	0.005	&	0.8874	&	9.7	\\
320	&	1.3373	&	0.005	&	0.4933	&	7.1	\\
321	&	0.6556	&	0.004	&	0.4132	&	6.5	\\
322	&	2.2856	&	0.005	&	0.4427	&	8.1	\\
323	&	1.4763	&	0.005	&	0.9978	&	7.2	\\
324	&	1.1644	&	0.005	&	0.8101	&	7.0	\\
325	&	1.1924	&	0.004	&	0.0803	&	7.2	\\
326	&	1.2013	&	0.005	&	0.5915	&	7.5	\\
327	&	0.2462	&	0.005	&	0.7426	&	5.9	\\
328	&	0.1468	&	0.005	&	0.8532	&	5.9	\\
329	&	0.1618	&	0.005	&	0.2239	&	5.8	\\
330	&	0.0583	&	0.007	&	0.2740	&	6.2	\\
331	&	0.5860	&	0.004	&	0.5058	&	7.0	\\
332	&	0.6462	&	0.006	&	0.1790	&	6.4	\\
333	&	1.5799	&	0.005	&	0.3919	&	7.3	\\
334	&	2.4147	&	0.005	&	0.8363	&	8.4	\\
335	&	3.0223	&	0.005	&	0.4275	&	9.0	\\
336	&	3.6313	&	0.006	&	0.2562	&	9.5	\\
337	&	0.1632	&	0.004	&	0.8586	&	5.8	\\
338	&	3.7341	&	0.005	&	0.4397	&	9.5	\\
339	&	1.8209	&	0.005	&	0.2187	&	7.6	\\
340	&	1.8070	&	0.005	&	0.6261	&	7.7	\\
341	&	0.1010	&	0.007	&	0.7478	&	5.7	\\
342	&	0.0763	&	0.006	&	0.2092	&	5.7	\\
343	&	0.0330	&	0.002	&	0.8370	&	6.3	\\
344	&	0.0377	&	0.007	&	0.2750	&	6.2	\\
345	&	0.1044	&	0.005	&	0.3752	&	5.8	\\
346	&	0.6069	&	0.006	&	0.4180	&	6.9	\\
347	&	0.6444	&	0.008	&	0.2527	&	6.5	\\
348	&	0.0032	&	0.007	&	0.6035	&	6.7	\\
349	&	0.0710	&	0.007	&	0.0744	&	6.0	\\
350	&	1.2278	&	0.005	&	0.1573	&	7.3	\\
351	&	1.5472	&	0.005	&	0.6609	&	7.2	\\
352	&	4.2246	&	0.005	&	0.6882	&	10.2	\\
353	&	4.9421	&	0.005	&	0.9629	&	10.7	\\
354	&	4.2356	&	0.005	&	0.0066	&	10.0	\\
355	&	1.6815	&	0.005	&	0.2665	&	7.3	\\
356	&	0.0746	&	0.005	&	0.9862	&	6.2	\\
357	&	0.0954	&	0.006	&	0.5252	&	6.0	\\
358	&	0.1461	&	0.004	&	0.0825	&	5.9	\\
359	&	0.1349	&	0.006	&	0.1508	&	5.6	\\
360	&	0.1538	&	0.005	&	0.2414	&	6.8	\\
361	&	0.1685	&	0.003	&	0.3625	&	5.9	\\
362	&	3.0186	&	0.005	&	0.4121	&	8.6	\\
363	&	3.0253	&	0.005	&	0.3291	&	9.0	\\
364	&	3.0164	&	0.005	&	0.1152	&	8.6	\\
365	&	2.4682	&	0.006	&	0.6312	&	9.4	\\
366	&	2.6966	&	0.005	&	0.9016	&	8.0	\\
367	&	0.6939	&	0.006	&	0.4330	&	6.2	\\
368	&	0.7096	&	0.005	&	0.8726	&	6.4	\\
369	&	0.6475	&	0.005	&	0.5853	&	6.5	\\
370	&	0.4486	&	0.004	&	0.3846	&	6.0	\\
371	&	0.1957	&	0.005	&	0.3543	&	5.6	\\
372	&	0.0682	&	0.007	&	0.6588	&	5.4	\\
373	&	0.4859	&	0.005	&	0.9815	&	6.0	\\
374	&	0.5481	&	0.004	&	0.8305	&	6.0	\\
375	&	0.4665	&	0.004	&	0.0094	&	5.9	\\
\end{tabular}

\begin{tabular}{cccccccccccccccccccc}
\hline\noalign{\smallskip}
N	&	F	&	A	&	P	&	SNR	\\
\hline\noalign{\smallskip}
376	&	4.2807	&	0.004	&	0.0670	&	9.6	\\
377	&	1.3744	&	0.004	&	0.5523	&	6.7	\\
378	&	2.3970	&	0.004	&	0.0194	&	7.6	\\
379	&	0.5398	&	0.005	&	0.0810	&	5.9	\\
380	&	0.6700	&	0.004	&	0.3677	&	6.0	\\
381	&	0.7127	&	0.004	&	0.9816	&	6.2	\\
382	&	0.6172	&	0.007	&	0.1135	&	7.0	\\
383	&	0.6520	&	0.005	&	0.4788	&	6.0	\\
384	&	3.2654	&	0.004	&	0.9935	&	8.2	\\
385	&	1.2430	&	0.004	&	0.7093	&	6.4	\\
386	&	0.0962	&	0.008	&	0.7486	&	5.9	\\
387	&	0.1095	&	0.004	&	0.6922	&	5.2	\\
388	&	2.4161	&	0.004	&	0.9073	&	7.4	\\
389	&	0.4732	&	0.003	&	0.0929	&	5.6	\\
390	&	0.6307	&	0.006	&	0.1380	&	6.0	\\
391	&	0.6096	&	0.008	&	0.1899	&	9.0	\\
392	&	0.6119	&	0.007	&	0.5343	&	7.2	\\
393	&	0.0780	&	0.004	&	0.7264	&	5.4	\\
394	&	1.0790	&	0.004	&	0.6642	&	6.2	\\
395	&	3.4950	&	0.004	&	0.1871	&	8.2	\\
396	&	1.4275	&	0.004	&	0.7428	&	6.4	\\
397	&	1.1896	&	0.005	&	0.6146	&	6.3	\\
398	&	1.2056	&	0.007	&	0.1518	&	8.3	\\
399	&	0.0242	&	0.006	&	0.5848	&	5.7	\\
400	&	0.0770	&	0.004	&	0.6703	&	5.9	\\
401	&	0.0135	&	0.004	&	0.5155	&	5.7	\\
402	&	0.0463	&	0.005	&	0.7925	&	6.2	\\
403	&	2.5190	&	0.004	&	0.9394	&	7.3	\\
404	&	4.7032	&	0.004	&	0.3849	&	9.5	\\
405	&	4.3381	&	0.004	&	0.6858	&	9.1	\\
406	&	1.8194	&	0.004	&	0.3103	&	7.1	\\
407	&	0.5656	&	0.006	&	0.0322	&	5.8	\\
408	&	2.7636	&	0.004	&	0.0579	&	7.5	\\
409	&	0.6157	&	0.006	&	0.1911	&	6.9	\\
410	&	0.5847	&	0.006	&	0.4052	&	6.1	\\
411	&	0.2017	&	0.005	&	0.3462	&	5.3	\\
412	&	3.1300	&	0.004	&	0.3363	&	7.8	\\
413	&	3.6218	&	0.004	&	0.9463	&	9.4	\\
414	&	0.1365	&	0.005	&	0.1730	&	5.2	\\
415	&	0.0390	&	0.007	&	0.8282	&	5.7	\\
416	&	0.0361	&	0.004	&	0.4996	&	6.2	\\
417	&	0.0617	&	0.005	&	0.8663	&	5.7	\\
418	&	0.1883	&	0.004	&	0.8337	&	5.5	\\
419	&	1.2639	&	0.004	&	0.5090	&	6.2	\\
420	&	0.6332	&	0.005	&	0.4889	&	6.3	\\
421	&	0.6255	&	0.006	&	0.7199	&	7.0	\\
422	&	0.5789	&	0.005	&	0.9912	&	6.3	\\
423	&	0.7763	&	0.004	&	0.4223	&	5.9	\\
424	&	2.7876	&	0.004	&	0.8879	&	7.4	\\
425	&	4.2334	&	0.005	&	0.4100	&	9.3	\\
426	&	0.1212	&	0.005	&	0.7227	&	5.6	\\
427	&	0.1320	&	0.003	&	0.9835	&	5.4	\\
428	&	0.6236	&	0.006	&	0.3066	&	6.2	\\
429	&	0.1128	&	0.005	&	0.7081	&	5.3	\\
430	&	3.0132	&	0.004	&	0.7375	&	7.7	\\
431	&	2.8910	&	0.004	&	0.2992	&	7.5	\\
432	&	3.8303	&	0.004	&	0.1565	&	8.4	\\
433	&	0.6917	&	0.005	&	0.3599	&	5.7	\\
434	&	0.5977	&	0.005	&	0.6593	&	6.4	\\
435	&	0.0149	&	0.004	&	0.3717	&	5.7	\\
436	&	4.8290	&	0.004	&	0.8955	&	9.6	\\
437	&	0.5458	&	0.003	&	0.7050	&	5.6	\\
438	&	0.5373	&	0.006	&	0.0530	&	5.8	\\
439	&	1.2251	&	0.004	&	0.0145	&	6.4	\\
440	&	0.8858	&	0.004	&	0.5837	&	5.8	\\
441	&	0.6359	&	0.006	&	0.8781	&	6.3	\\
442	&	2.5803	&	0.004	&	0.7516	&	7.0	\\
443	&	1.2408	&	0.005	&	0.6651	&	6.1	\\
444	&	2.8122	&	0.004	&	0.3518	&	7.3	\\
445	&	4.8352	&	0.005	&	0.3182	&	9.8	\\
446	&	1.8091	&	0.004	&	0.9382	&	6.6	\\
447	&	0.5834	&	0.007	&	0.3102	&	6.4	\\
448	&	0.5275	&	0.006	&	0.9830	&	5.6	\\
449	&	0.1141	&	0.005	&	0.0340	&	5.3	\\
450	&	0.0929	&	0.007	&	0.1916	&	5.7	\\
\end{tabular}
}
\end{center}
\end{table*}

\begin{table*}\tiny
\addtocounter{table}{-1}
\caption{Continued.}
\begin{center}
\resizebox{\textwidth}{!}{%
\begin{tabular}{cccccccccccccccccccc}
\hline\noalign{\smallskip}
N	&	F	&	A	&	P	&	SNR	\\
\hline\noalign{\smallskip}
451	&	0.2924	&	0.005	&	0.9228	&	5.1	\\
452	&	4.2316	&	0.005	&	0.9844	&	8.8	\\
453	&	4.2304	&	0.004	&	0.4969	&	9.4	\\
454	&	4.3576	&	0.004	&	0.8228	&	8.8	\\
455	&	1.8041	&	0.004	&	0.9325	&	6.7	\\
456	&	2.2187	&	0.004	&	0.9185	&	6.8	\\
457	&	4.8310	&	0.004	&	0.1935	&	9.3	\\
458	&	1.2346	&	0.004	&	0.5441	&	6.1	\\
459	&	1.8602	&	0.004	&	0.0884	&	6.3	\\
460	&	3.5830	&	0.004	&	0.8596	&	7.7	\\
461	&	0.6746	&	0.004	&	0.0340	&	5.4	\\
462	&	0.5685	&	0.005	&	0.8404	&	6.0	\\
463	&	0.5419	&	0.004	&	0.7589	&	5.6	\\
464	&	0.2362	&	0.004	&	0.7916	&	5.0	\\
465	&	0.1931	&	0.003	&	0.8427	&	5.0	\\
466	&	3.5256	&	0.004	&	0.5599	&	7.6	\\
467	&	1.1938	&	0.004	&	0.1118	&	5.9	\\
468	&	0.0705	&	0.007	&	0.8385	&	5.6	\\
469	&	0.6673	&	0.006	&	0.7820	&	5.6	\\
470	&	0.6685	&	0.005	&	0.2409	&	5.8	\\
471	&	0.4626	&	0.004	&	0.8015	&	5.1	\\
472	&	4.8361	&	0.004	&	0.9403	&	9.3	\\
473	&	4.9617	&	0.004	&	0.5742	&	8.9	\\
474	&	0.0945	&	0.006	&	0.3966	&	5.3	\\
475	&	1.9475	&	0.004	&	0.4560	&	6.5	\\
476	&	0.6667	&	0.006	&	0.1897	&	6.1	\\
477	&	0.6985	&	0.004	&	0.1118	&	5.4	\\
478	&	0.6803	&	0.004	&	0.6363	&	5.4	\\
479	&	0.0303	&	0.006	&	0.8319	&	5.3	\\
480	&	0.7038	&	0.004	&	0.1086	&	5.3	\\
481	&	0.3299	&	0.004	&	0.7127	&	4.9	\\
482	&	0.3033	&	0.004	&	0.6330	&	5.0	\\
483	&	0.1611	&	0.004	&	0.0336	&	4.9	\\
484	&	0.2035	&	0.004	&	0.1517	&	4.8	\\
485	&	0.2694	&	0.004	&	0.9749	&	4.8	\\
486	&	3.9009	&	0.004	&	0.1089	&	7.7	\\
487	&	0.1649	&	0.004	&	0.5565	&	4.9	\\
488	&	1.1698	&	0.003	&	0.7949	&	5.4	\\
489	&	1.2301	&	0.004	&	0.3618	&	5.6	\\
490	&	1.5136	&	0.003	&	0.8284	&	5.6	\\
491	&	0.5192	&	0.004	&	0.6906	&	4.9	\\
492	&	0.5221	&	0.003	&	0.0884	&	5.0	\\
493	&	0.3680	&	0.003	&	0.2712	&	4.8	\\
494	&	0.4213	&	0.003	&	0.8097	&	5.0	\\
495	&	0.1854	&	0.004	&	0.4820	&	4.7	\\
496	&	0.4602	&	0.003	&	0.5508	&	5.0	\\
497	&	0.2642	&	0.004	&	0.2734	&	4.6	\\
498	&	0.0652	&	0.008	&	0.0828	&	5.4	\\
499	&	0.0732	&	0.004	&	0.9467	&	4.8	\\
500	&	0.0555	&	0.007	&	0.3539	&	5.0	\\
501	&	0.1031	&	0.004	&	0.4012	&	5.0	\\
502	&	0.1074	&	0.005	&	0.6009	&	4.7	\\
503	&	4.0991	&	0.003	&	0.6757	&	7.6	\\
504	&	0.5535	&	0.003	&	0.4260	&	5.0	\\
505	&	2.8701	&	0.003	&	0.3971	&	6.6	\\
506	&	1.9436	&	0.004	&	0.2756	&	6.1	\\
507	&	4.8374	&	0.004	&	0.7261	&	9.0	\\
508	&	4.0504	&	0.003	&	0.7416	&	7.6	\\
509	&	3.0118	&	0.003	&	0.3390	&	6.7	\\
510	&	0.5441	&	0.003	&	0.0740	&	5.0	\\
511	&	0.4287	&	0.004	&	0.6468	&	4.8	\\
512	&	0.7063	&	0.003	&	0.3598	&	5.0	\\
513	&	2.4231	&	0.004	&	0.6310	&	6.4	\\
514	&	1.2313	&	0.004	&	0.4308	&	5.5	\\
515	&	1.8756	&	0.003	&	0.3359	&	5.8	\\
516	&	1.1806	&	0.004	&	0.1709	&	5.4	\\
517	&	0.1280	&	0.005	&	0.5063	&	4.8	\\
518	&	0.1297	&	0.005	&	0.3037	&	5.5	\\
519	&	0.1899	&	0.004	&	0.9423	&	4.8	\\
520	&	0.0457	&	0.005	&	0.1781	&	5.0	\\
521	&	2.0281	&	0.003	&	0.1735	&	5.9	\\
522	&	0.1244	&	0.005	&	0.9761	&	4.8	\\
523	&	0.1475	&	0.004	&	0.1391	&	4.5	\\
524	&	0.1595	&	0.004	&	0.5191	&	4.8	\\
525	&	0.5283	&	0.004	&	0.5056	&	5.0	\\
\end{tabular}
\begin{tabular}{cccccccccccccccccccc}
\hline\noalign{\smallskip}
N	&	F	&	A	&	P	&	SNR	\\
\hline\noalign{\smallskip}
526	&	0.5327	&	0.004	&	0.6999	&	4.9	\\
527	&	1.2648	&	0.004	&	0.9720	&	5.6	\\
528	&	1.2326	&	0.004	&	0.1543	&	5.5	\\
529	&	1.2165	&	0.006	&	0.9131	&	7.4	\\
530	&	1.2177	&	0.004	&	0.0758	&	6.4	\\
531	&	3.6302	&	0.003	&	0.3329	&	7.3	\\
532	&	1.0497	&	0.003	&	0.6360	&	5.1	\\
533	&	1.2232	&	0.004	&	0.6588	&	6.2	\\
534	&	1.2741	&	0.003	&	0.7266	&	5.4	\\
535	&	2.7576	&	0.003	&	0.1622	&	6.2	\\
536	&	0.1188	&	0.002	&	0.7721	&	4.8	\\
537	&	1.6829	&	0.003	&	0.8629	&	5.6	\\
538	&	4.1255	&	0.003	&	0.6723	&	7.3	\\
539	&	2.0233	&	0.004	&	0.9011	&	5.9	\\
540	&	2.0022	&	0.003	&	0.7112	&	5.8	\\
541	&	0.5383	&	0.004	&	0.3036	&	5.1	\\
542	&	1.2296	&	0.004	&	0.4024	&	6.5	\\
543	&	0.6535	&	0.005	&	0.5351	&	5.0	\\
544	&	0.4335	&	0.003	&	0.3702	&	4.7	\\
545	&	0.4434	&	0.003	&	0.2898	&	4.7	\\
546	&	0.4587	&	0.004	&	0.4281	&	4.8	\\
547	&	0.5246	&	0.003	&	0.5612	&	5.0	\\
548	&	1.1910	&	0.004	&	0.9239	&	5.3	\\
549	&	0.4798	&	0.004	&	0.5885	&	4.8	\\
550	&	0.0913	&	0.005	&	0.3688	&	4.6	\\
551	&	0.0795	&	0.006	&	0.8252	&	5.0	\\
552	&	0.1801	&	0.003	&	0.0581	&	4.7	\\
553	&	1.4486	&	0.003	&	0.7212	&	5.3	\\
554	&	2.3145	&	0.003	&	0.8615	&	5.9	\\
555	&	0.1409	&	0.004	&	0.2184	&	4.5	\\
556	&	0.1437	&	0.004	&	0.1211	&	4.8	\\
557	&	0.1703	&	0.003	&	0.8933	&	4.6	\\
558	&	0.0870	&	0.006	&	0.5919	&	5.0	\\
559	&	0.0228	&	0.004	&	0.8220	&	4.8	\\
560	&	2.5159	&	0.003	&	0.2398	&	5.9	\\
561	&	1.6358	&	0.003	&	0.9488	&	5.4	\\
562	&	0.2373	&	0.004	&	0.6719	&	4.7	\\
563	&	0.3482	&	0.003	&	0.6624	&	4.6	\\
564	&	2.4675	&	0.003	&	0.4860	&	6.1	\\
565	&	0.4299	&	0.004	&	0.1855	&	4.6	\\
566	&	0.1220	&	0.003	&	0.7294	&	4.6	\\
567	&	0.6679	&	0.005	&	0.4403	&	5.4	\\
568	&	0.5514	&	0.003	&	0.6977	&	4.8	\\
569	&	0.1510	&	0.003	&	0.6110	&	4.6	\\
570	&	0.0880	&	0.004	&	0.0630	&	5.1	\\
571	&	0.1727	&	0.004	&	0.9597	&	4.6	\\
572	&	0.0193	&	0.005	&	0.0172	&	5.2	\\
573	&	1.2469	&	0.003	&	0.1794	&	5.2	\\
574	&	2.4113	&	0.003	&	0.7440	&	5.9	\\
575	&	0.6451	&	0.003	&	0.6764	&	5.1	\\
576	&	0.7395	&	0.004	&	0.9102	&	4.8	\\
577	&	0.7075	&	0.003	&	0.6798	&	4.9	\\
578	&	1.8285	&	0.003	&	0.3248	&	5.5	\\
579	&	0.7002	&	0.003	&	0.8695	&	4.8	\\
580	&	0.7624	&	0.003	&	0.9665	&	4.8	\\
581	&	0.6500	&	0.003	&	0.8338	&	5.1	\\
582	&	0.5005	&	0.003	&	0.4009	&	4.7	\\
583	&	0.5097	&	0.004	&	0.7355	&	4.7	\\
584	&	0.1838	&	0.004	&	0.3237	&	4.4	\\
585	&	4.2229	&	0.002	&	0.0815	&	7.1	\\
586	&	0.2413	&	0.003	&	0.9250	&	4.3	\\
587	&	3.2706	&	0.003	&	0.4784	&	6.2	\\
588	&	0.6871	&	0.003	&	0.5815	&	4.7	\\
589	&	1.9979	&	0.003	&	0.9963	&	5.6	\\
590	&	0.2699	&	0.003	&	0.8092	&	4.7	\\
591	&	0.0172	&	0.005	&	0.0429	&	4.9	\\
592	&	3.6189	&	0.003	&	0.2065	&	6.4	\\
593	&	3.6282	&	0.003	&	0.9859	&	7.2	\\
594	&	3.6195	&	0.003	&	0.5458	&	6.7	\\
595	&	1.1290	&	0.003	&	0.3133	&	4.9	\\
596	&	2.2870	&	0.003	&	0.4770	&	5.6	\\
597	&	1.1522	&	0.003	&	0.4301	&	4.9	\\
598	&	4.2264	&	0.003	&	0.4045	&	7.1	\\
599	&	0.8963	&	0.003	&	0.7366	&	4.7	\\
600	&	0.4467	&	0.003	&	0.6903	&	4.5	\\
\end{tabular}
\begin{tabular}{cccccccccccccccccccc}
\hline\noalign{\smallskip}
N	&	F	&	A	&	P	&	SNR	\\
\hline\noalign{\smallskip}
601	&	0.4499	&	0.003	&	0.9200	&	4.6	\\
602	&	0.5465	&	0.004	&	0.9503	&	4.8	\\
603	&	0.5610	&	0.004	&	0.6445	&	4.9	\\
604	&	0.6393	&	0.004	&	0.5014	&	5.1	\\
605	&	0.5499	&	0.004	&	0.6060	&	4.8	\\
606	&	0.5237	&	0.004	&	0.2401	&	5.1	\\
607	&	0.8415	&	0.003	&	0.1400	&	4.7	\\
608	&	0.1488	&	0.003	&	0.3449	&	4.6	\\
609	&	0.0056	&	0.005	&	0.1989	&	5.5	\\
610	&	0.1081	&	0.004	&	0.6553	&	4.8	\\
611	&	1.8214	&	0.003	&	0.4326	&	6.3	\\
612	&	0.0407	&	0.005	&	0.0419	&	4.9	\\
613	&	0.0276	&	0.004	&	0.0953	&	5.5	\\
614	&	0.1749	&	0.004	&	0.7462	&	4.4	\\
615	&	0.1999	&	0.004	&	0.2315	&	4.5	\\
616	&	3.8855	&	0.003	&	0.1700	&	6.4	\\
617	&	0.4890	&	0.003	&	0.9331	&	4.6	\\
618	&	0.5404	&	0.004	&	0.2466	&	5.2	\\
619	&	0.7427	&	0.003	&	0.1378	&	4.6	\\
620	&	0.5490	&	0.003	&	0.4207	&	5.1	\\
621	&	2.5877	&	0.003	&	0.1844	&	5.6	\\
622	&	2.7741	&	0.003	&	0.7614	&	5.6	\\
623	&	1.2389	&	0.003	&	0.2676	&	4.9	\\
624	&	0.7211	&	0.003	&	0.4624	&	4.6	\\
625	&	0.7292	&	0.003	&	0.0699	&	4.6	\\
626	&	0.5024	&	0.003	&	0.4594	&	4.5	\\
627	&	0.7781	&	0.003	&	0.1803	&	4.6	\\
628	&	0.7586	&	0.003	&	0.1115	&	4.6	\\
629	&	2.0355	&	0.003	&	0.9931	&	5.3	\\
630	&	4.6816	&	0.003	&	0.5198	&	6.9	\\
631	&	0.7543	&	0.003	&	0.4304	&	4.5	\\
632	&	0.7321	&	0.003	&	0.9189	&	4.6	\\
633	&	1.8888	&	0.003	&	0.5241	&	5.1	\\
634	&	3.0178	&	0.003	&	0.8478	&	5.9	\\
635	&	2.5127	&	0.003	&	0.0020	&	5.5	\\
636	&	0.3277	&	0.003	&	0.4628	&	4.3	\\
637	&	1.3621	&	0.003	&	0.6154	&	4.8	\\
638	&	0.5667	&	0.004	&	0.1746	&	5.0	\\
639	&	4.8276	&	0.003	&	0.3592	&	7.0	\\
640	&	1.9965	&	0.003	&	0.7240	&	5.2	\\
641	&	2.0227	&	0.003	&	0.3445	&	5.4	\\
642	&	3.0476	&	0.003	&	0.8341	&	5.7	\\
643	&	1.8024	&	0.003	&	0.1989	&	5.2	\\
644	&	0.0853	&	0.003	&	0.1046	&	4.4	\\
645	&	0.4362	&	0.003	&	0.3771	&	4.4	\\
646	&	0.9193	&	0.003	&	0.7840	&	4.5	\\
647	&	0.5041	&	0.003	&	0.7596	&	4.6	\\
648	&	0.3401	&	0.003	&	0.7471	&	4.2	\\
649	&	0.8521	&	0.003	&	0.9344	&	4.5	\\
650	&	2.4583	&	0.003	&	0.2607	&	5.3	\\
651	&	1.2073	&	0.004	&	0.7362	&	6.0	\\
652	&	1.2888	&	0.003	&	0.1126	&	4.8	\\
653	&	1.1932	&	0.003	&	0.8973	&	4.9	\\
654	&	1.2812	&	0.003	&	0.9554	&	4.8	\\
655	&	1.2681	&	0.003	&	0.2097	&	4.8	\\
656	&	0.2740	&	0.003	&	0.1674	&	4.2	\\
657	&	0.2670	&	0.003	&	0.7951	&	4.3	\\
658	&	0.3104	&	0.003	&	0.4277	&	4.3	\\
659	&	0.3823	&	0.003	&	0.4068	&	4.3	\\
660	&	0.3237	&	0.004	&	0.9759	&	4.3	\\
661	&	0.3218	&	0.003	&	0.3380	&	4.4	\\
662	&	0.2936	&	0.003	&	0.3850	&	4.3	\\
663	&	0.2917	&	0.003	&	0.0415	&	4.4	\\
664	&	0.6267	&	0.004	&	0.7148	&	6.0	\\
665	&	0.0980	&	0.003	&	0.8208	&	4.9	\\
666	&	0.2730	&	0.003	&	0.6144	&	4.6	\\
667	&	0.5753	&	0.003	&	0.1982	&	4.8	\\
668	&	2.4105	&	0.003	&	0.8714	&	5.4	\\
669	&	3.7541	&	0.003	&	0.6437	&	7.1	\\
670	&	3.2042	&	0.003	&	0.1118	&	5.5	\\
671	&	3.0146	&	0.003	&	0.5768	&	5.7	\\
672	&	3.0274	&	0.003	&	0.4952	&	5.7	\\
673	&	3.0283	&	0.003	&	0.5680	&	6.0	\\
674	&	4.5341	&	0.003	&	0.3938	&	6.5	\\
675	&	1.3341	&	0.003	&	0.6482	&	4.9	\\
\end{tabular}
}
\end{center}
\end{table*}

\begin{table*}\scriptsize
\addtocounter{table}{-1}
\caption{Continued.}
\begin{center}
\begin{tabular}{cccccccccccccccccccc}
\hline\noalign{\smallskip}
N	&	F	&	A	&	P	&	SNR	\\
\hline\noalign{\smallskip}
676	&	0.3005	&	0.003	&	0.5771	&	4.2	\\
677	&	1.2216	&	0.003	&	0.0587	&	5.2	\\
678	&	3.9971	&	0.003	&	0.4223	&	6.1	\\
679	&	1.8259	&	0.003	&	0.3384	&	5.0	\\
680	&	1.8076	&	0.003	&	0.8685	&	5.2	\\
681	&	2.8270	&	0.003	&	0.5419	&	5.4	\\
682	&	0.4640	&	0.002	&	0.4254	&	4.3	\\
683	&	0.4567	&	0.003	&	0.5376	&	4.4	\\
684	&	0.0664	&	0.004	&	0.5410	&	4.8	\\
685	&	4.8297	&	0.003	&	0.6131	&	6.8	\\
686	&	0.0549	&	0.004	&	0.5179	&	4.8	\\
687	&	0.0425	&	0.004	&	0.6134	&	4.5	\\
688	&	0.3584	&	0.003	&	0.0050	&	4.3	\\
689	&	0.1907	&	0.003	&	0.9363	&	4.6	\\
690	&	0.1162	&	0.003	&	0.8436	&	4.8	\\
691	&	0.3911	&	0.003	&	0.3221	&	4.3	\\
692	&	0.6542	&	0.003	&	0.7953	&	4.6	\\
693	&	0.1395	&	0.003	&	0.6188	&	4.3	\\
694	&	0.2577	&	0.003	&	0.0831	&	4.2	\\
695	&	0.1088	&	0.003	&	0.7113	&	4.6	\\
696	&	2.0526	&	0.003	&	0.3691	&	5.1	\\
697	&	1.9362	&	0.003	&	0.8072	&	5.0	\\
698	&	1.3772	&	0.003	&	0.8655	&	4.7	\\
699	&	2.4126	&	0.003	&	0.1311	&	5.4	\\
700	&	0.6570	&	0.003	&	0.2109	&	4.9	\\
701	&	0.5151	&	0.003	&	0.4084	&	4.4	\\
702	&	0.8439	&	0.003	&	0.7323	&	4.3	\\
703	&	1.0644	&	0.003	&	0.8552	&	4.5	\\
704	&	1.8472	&	0.003	&	0.5657	&	4.9	\\
705	&	4.2238	&	0.003	&	0.9221	&	6.5	\\
706	&	0.9428	&	0.003	&	0.3426	&	4.3	\\
707	&	0.6134	&	0.004	&	0.3359	&	5.1	\\
708	&	0.5909	&	0.003	&	0.1009	&	4.9	\\
709	&	0.3504	&	0.003	&	0.7099	&	4.0	\\
710	&	0.3447	&	0.003	&	0.9244	&	4.2	\\
711	&	0.6714	&	0.004	&	0.3836	&	4.5	\\
712	&	0.6435	&	0.003	&	0.1370	&	5.3	\\
713	&	0.5451	&	0.003	&	0.1752	&	4.8	\\
714	&	0.6725	&	0.003	&	0.7879	&	4.7	\\
715	&	0.6979	&	0.003	&	0.8107	&	4.7	\\
716	&	0.7168	&	0.003	&	0.7007	&	4.4	\\
717	&	0.6623	&	0.003	&	0.1877	&	4.9	\\
718	&	3.9300	&	0.003	&	0.8226	&	5.9	\\
719	&	1.2402	&	0.003	&	0.0255	&	4.7	\\
720	&	1.2586	&	0.003	&	0.6966	&	4.7	\\
721	&	0.5260	&	0.003	&	0.7856	&	4.5	\\
722	&	2.8341	&	0.002	&	0.3379	&	5.2	\\
723	&	0.2627	&	0.003	&	0.1238	&	4.1	\\
724	&	0.5121	&	0.003	&	0.0645	&	4.4	\\
725	&	0.7406	&	0.003	&	0.0371	&	4.5	\\
726	&	0.5621	&	0.003	&	0.6052	&	4.7	\\
727	&	0.4930	&	0.003	&	0.8863	&	4.3	\\
728	&	1.8162	&	0.003	&	0.0778	&	5.1	\\
729	&	0.8539	&	0.002	&	0.1619	&	4.3	\\
730	&	1.2447	&	0.003	&	0.8448	&	4.7	\\
731	&	3.9228	&	0.002	&	0.3068	&	5.7	\\
732	&	1.3589	&	0.002	&	0.2463	&	4.5	\\
733	&	3.2856	&	0.002	&	0.2462	&	5.3	\\
734	&	3.8896	&	0.002	&	0.5791	&	5.7	\\
735	&	0.2824	&	0.002	&	0.4703	&	4.0	\\
\noalign{\smallskip}\hline
\end{tabular}
\end{center}
\end{table*}

\end{appendix}



\end{document}